%% file: 6csKz_fin.tex
\documentclass{article}
\usepackage{kmn,latexsym,subfigure,epsfig,graphics}
\newcommand{\mc}{\multicolumn}

\input macro.tex

\def \kband{$K$-band }
\def \rband{$R$-band }
\def \iband{$I$-band }
\def \ir1{IRCAM}
\def \irc3{IRCAM-3}

\def \uf{UFTI}

\def \ukirt{United Kingdom Infrared Telescope (UKIRT)}
\def \cosone{$\Omega_{\rm M}=1$ and $\Omega_{\Lambda}=0$\,}
\def \costwo{$\Omega_{\rm M}=0.3$ and $\Omega_{\Lambda}=0.7$\,}

\begin{document}

\title[The 6C* sample III: imaging data]{A sample of 6C radio sources
designed to find objects at redshift $> 4$: III --- imaging
and the radio galaxy $K-z$ relation}

\author[Jarvis et al.]
{Matt J.\,Jarvis$^{1,2}$\thanks{Email:jarvis@strw.leidenuniv.nl}, Steve Rawlings$^{1}$, Steve Eales$^{3}$,
\and
Katherine M.\,Blundell$^{1}$, Andrew
J.\,Bunker$^{4,5}$, Steve Croft$^{1}$,
\and 
Ross J.\,McLure$^{1}$, Chris J.\,Willott$^{1}$ \\
$^{1}$Astrophysics, Department of Physics, Keble Road, Oxford, OX1 3RH, U.K. \\
$^{2}$Sterrewacht Leiden, Postbus 9513, 2300 RA Leiden, the
Netherlands \\
$^{3}$Department of Physics and Astronomy, University of Wales 
College of Cardiff, P.O. Box 913, Cardiff, CF2 3YB, UK\\
$^{4}$Institute of Astronomy, University of Cambridge, Madingley Road,
Cambridge CB3 0HA, U.K. \\
$^{5}$Astronomy Department, University of California at Berkeley, CA 94720\\
}
\maketitle

\begin{abstract}

In this paper, the third and final of a series, we present complete
$K-$ band imaging and some complementary $I-$band imaging of the
filtered 6C* sample. We find no systematic differences between the
$K-z$ relation of 6C* radio galaxies and those from complete samples,
so the near-infrared properties of luminous radio galaxies are not
obviously biased by the additional 6C* radio selection criteria (steep
spectral index and small angular size). The 6C* $K-z$ data
significantly improve delineation of the $K-z$ relation for radio
galaxies at high-redshift ($z >2$). Accounting for non-stellar
contamination, and for correlations between radio luminosity and
stellar mass, we find little support for previous claims that the
underlying scatter in the stellar luminosity of radio galaxies
increases significantly at $z >2$. In a particular spatially-flat
universe with a cosmological constant (\costwo), the most luminous
radio sources appear to be associated with galaxies with a luminosity
distribution with a high mean ($\approx 5 L^{\star}$), and a low
dispersion ($\sigma \sim 0.5$\,mag) which formed their stars at epochs
corresponding to $z\,\gtsim\,2.5$. This result is in line with recent
sub-mm studies of high-redshift radio galaxies and the inferred ages
of extremely red objects from faint radio samples.
\end{abstract}


\section{Introduction}\label{sec:kzintro}
The powerful radio galaxies and quasars we see in the local universe
are almost exclusively associated with giant elliptical
galaxies. Studies of these objects at high redshift have shown that
the relationship between the near-infrared \kband and redshift $z$
(known as the $K-z$ diagram) is tight for the members of the 3CRR
sample (Laing, Riley \& Longair 1983) out to redshifts of $z \sim 2$.
The tightness of this correlation was initially used to assert that
the giant ellipticals hosting the low-redshift ($z < 0.6$) radio
galaxies from the 3CRR sample are the result of passively-evolving
stellar populations which formed at high-redshift (e.g. Lilly 1989).

Frequently, at higher redshifts ($z > 0.6$) radio galaxies exhibit
optical/infrared emission aligned with their radio axes. This aligned
emission is not expected if the host galaxy emission is a consequence
of an old stellar population alone. The mechanism which causes this
`alignment effect' has been the subject of much debate over the past
decade. Different possibilities include, incorporating jet driven star bursts
(e.g. McCarthy et al. 1987; Chambers, Miley \& van Breugel 1987),
scattered quasar light (e.g. Tadhunter et al. 1987) and emission from
inverse Compton scattering of the cosmic microwave background
(e.g. Daly et al. 1992) have all been put forward as possible
explanations. A combination of the three mechanisms mentioned above,
as well as others, may explain the alignment effect in the majority of
sources.

Dunlop \& Peacock (1993) using 3CRR sources along with fainter Parkes
radio galaxies found that this alignment effect is weaker in the
near-infrared than in the optical, indicating that the aligned
component is blue in colour. They also found that the alignment was a
strong function of radio luminosity at $z \sim 1$, with the most
luminous sources exhibiting a greater degree of alignment.

Best, Longair \& R\"ottgering (1997, 1998) and McLure \& Dunlop (2000)
both find that the hosts of 3CRR radio galaxies at $z \sim 1$ are
consistent with massive evolved ellipticals, whose infrared emission
is dominated by starlight. McLure \& Dunlop also demonstrate that the
hosts of 3CRR radio galaxies at $z \sim 1$ are essentially the same as
the hosts of 3CRR radio galaxies at $z \sim 0.2$. The small amount of
passive evolution between $z \sim 1$ and $z \sim 0.2$ implies a
high-formation redshift ($z > 3$), within most reasonable cosmologies,
for these massive elliptical host galaxies.

Eales et al. (1997) using the 3CRR sample and the fainter 6CE sample
(Eales 1985; Rawlings, Eales \& Lacy 2001) found that lower luminosity radio
sources are associated with optically fainter host galaxies. This led
Eales et al. (1997) to conclude that a correlation exists between the
radio luminosity and optical/infrared luminosity.

Another factor which is dependent on radio luminosity and
contributes to the optical/infrared emission is the contamination from
bright emission lines which have been redshifted into the \kband
window (e.g. Eales \& Rawlings 1993). This contamination from emission
lines means that the stellar population is not the only contributor to
the \kband flux. It is now well documented that the emission-line
strength in radio galaxies is intrinsically linked to the strength of
the underlying quasar continuum and thus the radio emission (e.g. Rawlings \&
Saunders 1991; Willott et al. 1999; Jarvis et al. 2001). Therefore,
due to the biases inherent in a single flux-density-limited sample
such as 3CRR (e.g. Blundell, Rawlings \& Willott 1999), the most
distant sources are always the most luminous, which in turn produce
the strongest emission lines. At low-redshift this contamination is
negligible. However, at $z > 1$ the effect becomes increasingly
important, as the bright optical emission lines are redshifted into
the near-infrared. With fainter samples such as 6CE, this effect has,
to a certain degree been quashed due to the low-luminosities compared
to 3CRR. However, at $z > 2$ the luminosities of these sources are
similar to the 3CRR sources at $z \sim 0.5$, thus the emission-line
contributions need to be considered. These effects may thus have an
important effect on the $K-z$ diagram at high redshift.

Eales et al. (1997) also conclude that at redshifts $z > 2$ we are beginning to see
the formation epoch of these massive ellipticals. The relative
brightness of the 6CE sources at $z > 2$ in comparison to a
passively-evolving stellar population, and also the higher dispersion in the
\kband magnitudes at high redshift both point to this
explanation. However, the relatively low number of sources above $z >
2$ in the 6CE sample place considerable uncertainties on this
high-redshift behaviour.

In this paper we present our \kband imaging data for all sources in
the 6C* sample.  The 6C* sample is a filtered sample, aimed at finding
radio galaxies at $z >4$. The filtering criteria employed are based on
the radio properties of the sources, namely steep-spectral index
($\alpha \ge 0.981$\footnote{We use the convention for spectral index
$S_{\nu} \propto \nu^{-\alpha}$, where $\alpha$ is measured between
observed 151~MHz and 4.9~GHz.}) and small angular size ($\theta <
15$\,arcsec). A detailed description of the sample selection and
filtering criteria is given in Blundell et al. (1998; paper I). The
spectrophotometric observations are presented in (Jarvis et al. 2001;
paper II).  We
are now able to use this filtered sample, along with the 6CE sample
(Eales 1985; Rawlings et al. 2001), and the high-redshift sources from the
7C-III sample of Lacy, Bunker \& Ridgway (2000) to investigate the
stellar continuum of high-redshift radio galaxies with the $K-z$
Hubble diagram up to $z \approx 4.4$.

In Sec.~\ref{sec:kband_imaging} we outline the observing and reduction
procedures for our \kband imaging data. We also present \kband images
for 28 of the 29 sources in the final 6C* sample with their radio
contours overlaid. A full discussion and images of the other member of
the 6C* sample, 6C*0140+326, which is at $z = 4.41$ may be found in
Rawlings et al. (1996). In Sec.~\ref{sec:notes} we provide brief notes of
each source in the sample. We investigate the shape and properties of
the $K-z$ Hubble diagram to $z \approx 4.4$ in Sec.~\ref{sec:kz}, and in
Sec.~\ref{sec:mainevolution} we suggest an evolutionary scenario to
explain this diagram. We summarise our conclusions in
Sec.~\ref{sec:kzconclusions}. We also present $I-$band images
of some of the 6C* sources in Appendix A, and imaging data on the
sources no longer deemed to be members of 6C* in Appendix B. For
consistency with previous papers we use $H_{\circ} = 50\,{\rm
km\,s}^{-1}\,{\rm Mpc}^{-1}$, in a spatially flat cosmology with
\cosone, unless otherwise stated.

\section{\kband imaging}\label{sec:kband_imaging}
\subsection{Observations}
The \kband (2.2$\mu$m) imaging of the 6C* sample was made over several
observing runs at the \ukirt, beginning in 1993, with the most recent
observations made in September 2000. The earliest observations (1993 -
1994) were all made using the \ir1 near-infrared camera, which
consists of a $58 \times 62$ element infrared array, with a pixel
scale of 0.62\,arcsec pixel$^{-1}$. For the second observing run in
1997, \irc3 was used. \irc3 is a InSb
$256 \times 256$ array with a pixel scale of 0.286 arcsec pixel$^{-1}$
and field of view $73 \times 73$ arcsec$^{2}$. The most recent
observations were made using the new infrared camera on UKIRT, UFTI
(UKIRT Fast-Track Imager), which comprises a $1024 \times 1024$ HgCdTe
array, with a plate scale of 0.091 arcsec pixel$^{-1}$, giving a field
of view of $92 \times 92$ arcsec$^{2}$. All observations were made in
photometric conditions unless stated otherwise, with the typical
seeing spanning 0.6 - 1.2 arcsec. From 1997 onwards, the tip-tilt
system on UKIRT was utilised to give sub-arcsec imaging on all
observations.
	
In order to subtract the rapidly changing sky background at these
wavelengths, to provide a good flat-field and to minimise the effects
of cosmic ray contamination and bad pixels, we used the standard
observing strategy of offsetting the position of the telescope by
e.g. 8\,arcsec between each exposure. The offsets were
arranged in a $3 \times 3$ mosaic of nine exposures. The integration time
varies between objects in the sample, due to time constraints and also
based on the magnitude of the source. A summary of all the \kband
observations is given in Table~\ref{tab:im_journalk}.

\subsection{Data reduction}
The \kband images were reduced using standard procedures. We first
subtracted the dark current from each image. We then divided by the
normalised flat-field, created by combining the nine exposures of the
particular field with a median filter, which removes any objects which
appear in different positions on the chip over the nine exposures. To
combine the individual images we registered all of the frames using a
bright star which was present in each of the nine pointings. In the cases where there
was no bright star in the images, the offsets recorded in the image
headers were used to align the frames. The registered images were then
combined using an average clipping procedure to reject pixels more
than 4$\sigma$ away from the median of the distribution. 

Astrometry for each image was achieved by identifying sources
in the image with objects in the APM catalogue\footnote{For more details see
http://www.ast.cam.ac.uk/$\sim$apmcat/}. In the few cases where
the APM did not cover the sky area required, ground-based images
with accurately determined astrometry were used, and in one case
(6C*0133+486) the Digitised Sky Survey (DSS) was used. In the majority
of the cases we were able to identify three or more sources on our
images with objects on the finding charts, and in these cases the
astrometry (performed with the IRAF task {\bf gasp}) is accurate to
$\ltsim\, 1$ arcsec. In the cases where three or more objects could
not be identified with objects on the finding charts, astrometric
scaling from the previous images was used to perform the astrometry on
the images from a single securely identified object.  The reduced
images are shown in Fig.~\ref{fig:kband_images} with the radio contours
overlaid.

\begin{table*}[!ht]
\begin{center}
\begin{tabular}{lrcr}
\hline\hline
\mc{1}{c}{Source} &
\mc{1}{c}{Instrument} & \mc{1}{c}{Date} & \mc{1}{c}{Exposure} \\
\mc{1}{c}{name} & & \mc{1}{c}{} & \mc{1}{c}{time / s} \\
\hline\hline
6C*0020+440  & {\bf \ir1} & {\bf 94Jan16}   &
{\bf 1440} \\
6C*0024+356  & {\bf \irc3} & {\bf 97Dec13}   &
{\bf 540} \\
6C*0031+403  &  {\bf \ir1} & {\bf 94Jan18}   &
{\bf 1260} \\
6C*0032+412    & {\bf \ir1} & {\bf 93Jan10}   &
{\bf 3240} \\	
6C*0041+469    & {\bf \ir1} & {\bf 94Jan16}   & {\bf
1440} \\
	     & \irc3 & 97Dec13 & 540 \\  
\hline
6C*0050+419    & \ir1 &  94Jan18 & 1260 \\
	     & {\bf \irc3} & {\bf 97Dec13} & {\bf 540} \\
             & \uf & 00Sep25 & 960 \\  
6C*0052+471    & {\bf \ir1} & {\bf 94Jan18}   & {\bf
630} \\
6C*0058+495    & {\bf \ir1} & {\bf 94Jan16}   &
{\bf 1440} \\
6C*0106+397 & \irc3 & 97Dec13 & 540 \\
               & {\bf \uf} & {\bf 00Sep25}   & {\bf
960} \\
6C*0112+372    & {\bf \uf} & {\bf 00Jul28}   & {\bf
540} \\
\hline
6C*0115+394    & {\bf \irc3} & {\bf 97Dec13}   &
{\bf 540} \\
6C*0118+486    & \ir1 & 94Jan16 & 1440 \\
               & {\bf \uf} & {\bf 00Jul28}   & {\bf
540} \\
6C*0122+426    & {\bf \irc3} & {\bf 97Dec13}   &
{\bf 540} \\
6C*0128+394    & {\bf \ir1} & {\bf 94Jan18}   &
{\bf 1260} \\
6C*0132+330    & {\bf \irc3} & {\bf 97Dec13}   &
{\bf 540} \\
\hline
6C*0133+486    & {\bf \irc3} & {\bf 97Dec13}   &
{\bf 540} \\
6C*0135+313    & {\bf \irc3} & {\bf 97Dec13}   &
{\bf 540} \\
6C*0136+388    & {\bf \ir1} & {\bf 94Jan16}   &
{\bf 1440} \\
6C*0139+344   & \ir1 & 94Jan18 & 630 \\
	       & {\bf \irc3} & {\bf 97Dec13}   &
{\bf 540} \\
6C*0142+427    & {\bf \ir1} & {\bf 94Jan16}   &
{\bf 1440} \\
\hline
6C*0152+463   & \ir1 & 94Jan16 & 1440 \\
               & {\bf \irc3} & {\bf 97Dec13}   &
{\bf 540} \\
6C*0154+450    & {\bf \irc3} & {\bf 97Dec13}   &
{\bf 540} \\
6C*0155+424   & \ir1 & 94Jan18 & 630 \\
	       & {\bf \irc3} & {\bf 97Dec13}   &
{\bf 540} \\
6C*0158+315   & \ir1 & 94Jan17 & 1260 \\
               & {\bf \irc3} & {\bf 97Dec13}   &
{\bf 540} \\
\hline
6C*0201+499    & {\bf \ir1} & {\bf 94Jan18}   &
{\bf 1260} \\
	       & \irc3 & 97Dec13 & 540 \\
6C*0202+478    & {\bf \ir1} & {\bf 94Jan18}   &
{\bf 1260} \\
6C*0208+344    & {\bf \ir1} & {\bf 94Jan17}   &
{\bf 1440} \\
	     & \irc3 & 97Dec13 & 540 \\
6C*0209+276    & {\bf \ir1} & {\bf 94Jan17}   &
{\bf 1620} \\
\hline\hline
\end{tabular}
\end{center}
{\caption{\label{tab:im_journalk} Log of the \kband imaging
observations of the sources present in the 6C$^{*}$ sample. The
imaging data on the $z = 4.41$ radio galaxy, 6C*0140+326 is presented
in Rawlings et al. (1996). The bold lettering indicates the images
shown in Fig.~\ref{fig:kband_images}. The Sep00 image of
6C*0050+419 is not of sufficient depth to detect the object in $K-$band,
due to non-photometric conditions, therefore we use the Dec97 image
for our analysis.}}
\end{table*}

\begin{table*}[!ht]
\small
\begin{center}
\begin{tabular}{llllllc}
\hline\hline
\mc{1}{c}{Source} & \mc{1}{c}{Magnitude from} & \mc{1}{c}{Magnitude from}
& \mc{1}{c}{Magnitude from} & \mc{1}{c}{$z$} & \mc{1}{l}{Detector} &
\mc{1}{l}{Radio}\\
\mc{1}{c}{} & \mc{1}{c}{3'' diameter} & \mc{1}{c}{5'' diameter}
& \mc{1}{c}{8'' diameter} & \mc{1}{c}{} & \mc{1}{l}{} &
\mc{1}{l}{frequency}\\
\hline\hline
6C*0020+440 & 19.94 $\pm$ 0.41 & 18.90 $\pm$ 0.27  & nbo &
2.988 & \ir1 & 8.9 GHz \\
6C*0024+356 & 18.51 $\pm$ 0.12 & nbo & nbo &  2.161 &
\irc3 & 8.4 GHz \\
6C*0031+403 & 17.68 $\pm$ 0.05 & 17.12 $\pm$ 0.06 & 16.80 $\pm$ 0.08 &
1.619 & \ir1 & 1.5 GHz \\
6C*0032+412 & 19.95 $\pm$ 0.43 & 19.62 $\pm$ 0.42 & 19.57 $\pm$ 0.55 & 3.658 & \ir1 & 4.9 GHz \\
6C*0041+469 & 19.36 $\pm$ 0.17 & 19.18 $\pm$ 0.27 & 19.19 $\pm$ 0.51 &
2.140? & \ir1 & 4.9 GHz \\
 & 20.13 $\pm$ 0.62 & 19.65 $\pm$ 0.73 & 19.32 $\pm$ 0.96 &
2.140 & \irc3 & \\
\hline
6C*0050+419 & 20.42 $\pm$ 0.53 & 20.39 $\pm$ 0.93 & 19.61 $\pm 1.35$ & 1.748? &
\ir1 & \\
 & 19.66 $\pm$ 0.37 & 19.43 $\pm$ 0.70 & 18.80 $\pm$ 0.54 &
1.748? & \irc3 & 8.4 GHz \\
6C*0052+471 & 18.01 $\pm$ 0.10 & 17.89 $\pm$ 0.15 & 17.83 $\pm$ 0.27 &
1.935 & \ir1 & 8.4 GHz \\
6C*0058+495 & 18.02 $\pm$ 0.06 & 17.73 $\pm$ 0.08 & 17.62 $\pm$ 0.14 &
1.173 & \ir1 & 4.9 GHz \\
6C*0106+397 & 19.25 $\pm$ 0.39 & 18.78 $\pm$ 0.36 & 18.43 $\pm$ 0.41
& 2.284 & \irc3 & 1.5 GHz \\
6C*0112+372 & 18.25 $\pm$ 0.05 & 18.12 $\pm$ 0.08 & 17.97 $\pm$ 0.13 &
2.535 & \uf & 4.9 GHz \\
\hline
6C*0115+394 & 18.88 $\pm$ 0.19 & 18.74 $\pm$ 0.30 & 18.58 $\pm$ 0.47 &
2.241 & \irc3 & 8.4 GHz \\
6C*0118+486 & 18.58 $\pm$ 0.09 & 18.28 $\pm$ 0.12 & 18.13 $\pm$ 0.20 &
2.350 & \ir1 & \\
& 18.49 $\pm$ 0.07 & 18.23 $\pm$ 0.10 & 18.31 $\pm$ 0.20 &
2.350 & \uf & 8.4 GHz \\
6C*0122+426 & 18.91 $\pm$ 0.19 & 18.88 $\pm$ 0.31 & 18.72 $\pm$ 0.35 & 2.635 &
\irc3 & 8.4 GHz \\
6C*0128+394 & 18.01 $\pm$ 0.07 & 17.78 $\pm$ 0.11 & 17.72 $\pm$ 0.22 &
0.929 & \ir1 & 8.4 GHz \\
6C*0132+330 & 19.30 $\pm$ 0.27 & 18.80 $\pm$ 0.29 & 18.81 $\pm$ 0.52 &
1.710 & \irc3 & 8.4 GHz \\
\hline
6C*0133+486 & 18.72 $\pm$ 0.16 & 18.69 $\pm$ 0.27 & nbo &
1.029? & \irc3 & 8.4 GHz \\ 
6C*0135+313 & 19.92 $\pm$ 0.59 & 19.47 $\pm$ 0.55 & nbo &
2.199 & \irc3 & 1.5 GHz \\
6C*0136+388 & 18.07 $\pm$ 0.05 & 17.57 $\pm$ 0.05 & 17.32 $\pm$ 0.08 &
1.108? & \ir1 & 4.9 GHz\\
6C*0139+344 & 18.17 $\pm$ 0.09 & nbo & nbo &
1.637 & \ir1 &  \\
& 18.40 $\pm$ 0.11 & nbo & nbo & 1.637 & \irc3 & 4.9 GHz \\
6C*0142+427 & 19.18 $\pm$ 0.17 & 18.81 $\pm$ 0.22 & 18.75 $\pm$ 0.53 &
2.225 & \ir1 & 1.5 GHz \\ 
\hline
6C*0152+463 & 19.04 $\pm$ 0.15 & 18.87 $\pm$ 0.23 & 18.99 $\pm$ 0.49 &
2.279 & \ir1 & \\
& 19.24 $\pm$ 0.27 & 18.55 $\pm$ 0.25 & 18.40 $\pm$ 0.39 &
2.279 & \irc3 & 4.9 GHz \\
6C*0154+450 & 17.06 $\pm$ 0.04 & 17.03 $\pm$ 0.06 & 16.99 $\pm$ 0.11 &
1.295 & \irc3 & 8.4 GHz \\
6C*0155+424S & 17.47 $\pm$ 0.05 & 17.17 $\pm$ 0.07 &  nbo  & 0.513 &
\ir1 &\\
& 17.62 $\pm$ 0.06 & 17.38 $\pm$ 0.08 &  nbo  & 0.513 & \irc3 & 8.4 GHz \\
6C*0158+315  & 18.65 $\pm$ 0.12 & 18.21 $\pm$ 0.14 &  17.91 $\pm$ 0.19
& 1.505 & \ir1 &  \\
& 18.82 $\pm$ 0.16 & 18.53 $\pm$ 0.21 & 18.17 $\pm$ 0.27 & 1.505 &
\irc3 & 4.9 GHz \\
\hline
6C*0201+499 & 18.76 $\pm$ 0.11 & 18.66 $\pm$ 0.18 & 18.50 $\pm$ 0.29 &
	1.981 & \ir1 & 4.9 GHz \\
& 19.27 $\pm$ 0.52 & 19.20 $\pm$0.42 & 18.68 $\pm$ 0.48 &
	1.981 & \irc3 & \\
6C*0202+478 & 18.90 $\pm$ 0.80 & nbo & nbo & 1.613? & \ir1 & 8.4 GHz \\
6C*0208+344 & 19.26 $\pm$ 0.21 & 19.03 $\pm$ 0.33 & 18.99 $\pm$ 0.95 &
1.920 & \ir1 & 4.9 GHz \\
& 19.62 $\pm$ 0.34 & 19.16 $\pm$ 0.41 & 19.05 $\pm$ 0.53 &
	1.920 & \irc3 &\\
6C*0209+276 & 18.66 $\pm$ 0.09 & 18.28 $\pm$ 0.12 & 18.10 $\pm$ 0.18 &
1.141? & \ir1 & 4.9 GHz \\
\hline\hline
\end{tabular}
\end{center}
{\caption{\label{tab:kmags} \kband magnitudes for the 6C* sample in
three different angular apertures. The
redshifts are the spectroscopic redshifts from Jarvis et al. (2001). `nbo' denotes
that the radio galaxy was too close to a nearby object to
measure the magnitude reliably. The \kband magnitude for 6C*0050+419
is for object `a' in Fig.~\ref{fig:kband_images}. The \kband magnitude of 6C*0139+344 may have a
contribution to the \kband flux from a foreground galaxy. Column 7
gives the frequency of the radio maps shown in Fig.~\ref{fig:kband_images}. The \kband magnitude for 6C*0202+478 is an estimate,
because of its close proximity to other nearby objects.}}
\end{table*}

\begin{table*}[!ht]
\small
\begin{center}
\begin{tabular}{lcccllllcll}
\hline\hline
\mc{1}{c|}{(1)} & \mc{1}{c|}{(2)} & \mc{1}{c|}{(3)} & \mc{1}{c|}{(4)} &
\mc{1}{c|}{(5)} & \mc{1}{c|}{(6)} & \mc{1}{c|}{(7)} & \mc{1}{c|}{(8)}
& \mc{1}{c|}{(9)} & \mc{1}{c|}{(10)} & \mc{1}{l|}{(11)}\\
\hline
\mc{1}{c|}{Source} & \mc{1}{c|}{$S_{151}$} &
\mc{1}{c|}{$\alpha_{151}$} & \mc{1}{c|}{$z$} & \mc{1}{c|}{$K$} &
\mc{1}{c|}{$R$} &
\mc{1}{c|}{$\log_{10}L_{151}$} & \mc{1}{c|}{Line} & 
\mc{1}{c|}{$\log_{10}L_{\rm line}$} & \mc{1}{l|}{Class} & \mc{1}{c|}{Ref}\\
\hline\hline
6C*0020+440 & 2.00 & 0.97 & 2.988 & 18.90 (5) & & 28.03 & Ly$\alpha$ &
36.37 & HEG & Jea \\
6C*0024+356 & 1.09 & 0.69 & 2.161 & 18.51 (3) & 22.80 & 27.33 & Ly$\alpha$ &
35.96 & LEG? & Jea  \\
6C*0031+403 & 0.96 & 0.90 & 1.618 & 16.80 (8) & 22.42 & 27.08 & CIV &
35.18 & HEG? & Jea \\
6C*0032+412 & 1.29 & 1.28 & 3.658 & 19.57 (8) & & 28.21 & Ly$\alpha$ &
36.35 & HEG & Jea \\
6C*0041+469 & 1.53 & 0.61 & 2.145? & 19.19 (8) & & 27.31 & Ly$\alpha$ &
35.55 & HEG? & Jea  \\
6C*0050+419 & 1.00 & 1.24 & 1.748? & 18.80 (8) & & 27.31 & HeII? & 34.16
& HEG? & Jea \\
\hline
6C*0052+471 & 1.31 & 0.88 & 1.935 & 17.83 (8) & & 27.38 & Ly$\alpha$ &
36.12 & Q & Jea  \\
6C*0058+495 & 0.97 & 0.74 & 1.173 & 17.62 (8) & & 26.74 & [OII] & 35.92
& HEG & Jea  \\
6C*0106+397 & 0.96 & 0.52 & 2.284 & 18.43 (8) & & 27.25 & CIV & 34.87
& HEG & Jea  \\
6C*0112+372 & 1.03 & 0.42 & 2.535 & 17.97 (8) & & 27.36 & Ly$\alpha$ &
37.05 & HEG & Jea \\
6C*0115+394 & 0.96 & 1.06 & 2.241 & 18.58 (8) & 22.31 & 27.48 & Ly$\alpha$ &
37.15 & HEG & Jea \\
6C*0118+486 & 0.98 & 1.47 & 2.350 & 18.13 (8) & & 27.70 & Ly$\alpha$ &
36.17 & HEG & Jea  \\
\hline
6C*0122+426 & 1.05 & 0.53 & 2.635 & 18.88 (5) & 23.10 & 27.43 & Ly$\alpha$ &
35.97 & HEG & Jea \\
6C*0128+394 & 1.94 & 0.50 & 0.929 & 17.72 (8) & & 26.75 & [OII] & 34.52
& HEG? & Jea \\
6C*0132+330 & 1.56 & 1.28 & 1.710 & 18.81 (8) & 22.12 & 27.50 & CIII] & 35.19
& HEG & Jea \\
6C*0133+486 & 1.89 & 1.22 & 1.029? & 18.69 (5) & 23.39 & 27.04 & [OII]? &
35.05 & LEG? & Jea \\
6C*0135+313 & 1.24 & 1.18 & 2.199 & 19.47 (5) & 23.82 & 27.62 & Ly$\alpha$ &
37.71 & HEG & Jea \\
6C*0136+388 & 0.99 & 0.68 & 1.108? & 17.32 (8) & 22.60 & 26.68 & [OII]? &
35.68 & LEG? & Jea \\
\hline
6C*0139+344 & 1.10 & 0.74 & 1.637 & 18.40 (3) & & 27.11 & [OII] & --- &
HEG & Jea \\
6C*0140+326 & 1.00 & 0.62 & 4.410 & 20.00$^{\dagger}$ (2) & & 27.94 &
Ly$\alpha$ & 36.87 & HEG & Rea  \\  
6C*0142+427 & 1.46 & 1.20 & 2.225 & 18.75 (8) & & 27.70 & Ly$\alpha$ &
36.72 & HEG & Jea  \\
6C*0152+463 & 1.29 & 0.79 & 2.279 & 18.40 (8) & & 27.49 & Ly$\alpha$ &
36.15 & LEG? & Jea \\
6C*0154+450 & 1.15 & 1.28 & 1.295 & 16.99 (8) & & 27.07 & CIV & 37.22
& Q & Jea \\
6C*0155+424 & 1.51 & 0.89 & 0.513 & 17.38 (5) & 20.76 & 26.20 & [OII] & 34.66
& LEG? & Jea \\
\hline
6C*0158+315 & 1.51 & 0.75 & 1.505 & 17.91 (8) & 22.28 & 27.16 & MgII & 36.31 &
HEG? & Jea \\
6C*0201+499 & 1.14 & 0.74 & 1.981 & 18.50 (8) & & 27.30 & Ly$\alpha$ &
35.75 & HEG & Jea \\
6C*0202+478 & 1.06 & 0.86 & 1.620? & 18.90$^{\ddagger}$ (3) & & 27.13
& CII]? & 36.10 & HEG? & Jea \\
6C*0208+344 & 0.97 & 0.57 & 1.920 & 19.03 (5) & & 27.13 & Ly$\alpha$ &
36.01 & HEG & Jea \\
6C*0209+476 & 1.14 & 0.69 & 1.141? & 18.10 (8) & & 26.77 & [OII]? &
35.18 & LEG? & Jea \\
\hline\hline
Excluded sources & & & & & & & & & \\
6C*0100+312 & 1.16 & 1.68 & 1.189 & 16.54 (8) & --- & 27.14 & MgII & 36.14
& Q & Jea \\
6C*0107+448 & 1.04 & 1.42 & 1.301? & --- & --- & 27.09 & [OII]? &
34.82 & G & Jea \\
6C*0111+367 & 1.11 & --- & --- & 17.80 (8) & --- & --- & --- & --- & G & Jea \\
6C*0120+329 & 1.87 & 1.48 & 0.0164 & --- & --- & 23.24 & --- & --- & G & deR \\
6C*0141+425 & 1.64 & 1.06 & 0.0508 & --- & --- & 24.16 & abs & --- & G & Jea \\ 
\hline\hline
\end{tabular}
\end{center}
{\caption{\label{tab:6cssummary} Summary of the observational data on
the 6C* sample (an ascii version of this table along with other
information about the 6C* sample can also be found at http://www.strw.leidenuniv.nl/$\sim$jarvis/6cs/).  {\bf Column 1:} Name of the 6C* source. {\bf Column
2:} 151 MHz flux-density measurements in Jy from the 6C survey (Hales
et al. 1993). {\bf Column 3:} radio spectral index evaluated at
rest-frame 151~MHz using the polynomial fit described in Blundell et
al. (1998). {\bf Column 4:} redshift, `?' signifies that this value is
uncertain. {\bf Column 5:} \kband magnitudes within the angular
aperture in arcseconds given in brackets. $\dagger$ signifies that the
measurement is from van Breugel et al. (1998) and is a $K^{\prime}$
magnitude, $\ddagger$ signifies that the optical ID is too close to a
nearby object to measure the \kband magnitude with any accuracy. {\bf
Column 6:} \rband magnitude where available (Jarvis 2000), measured
using the same aperture as that for the \kband magnitude. {\bf Column
7:} $\log_{10}$ of the rest-frame 151 MHz radio luminosity (measured
in units of W Hz$^{-1}$ sr$^{-1}$), calculated using the polynomial
fit to the radio spectra for $\Omega_{\rm M} = 1$,\,$\Omega_{\Lambda}
= 0$, $H_{\circ} = 50\,{\rm km\,s}^{-1}\,{\rm Mpc}^{-1}$. {\bf Column
8:} Prominent emission line in the existing spectra, `?' signifies
that the line identification is uncertain. {\bf Column 9:} $\log_{10}$
of the luminosity of the line listed in Column 8 (measured in units of
W), for $\Omega_{\rm M} = 1$,\,$\Omega_{\Lambda} = 0$ and $H_{\circ} =
50\,{\rm km\,s}^{-1}\,{\rm Mpc}^{-1}$.  `-' signifies that the data
were inadequate to obtain a line luminosity through the absence of a
spectrophotometric standard. {\bf Column 10:} classification,
Q=quasar, HEG=high-excitation galaxy, LEG=low-excitation galaxy,
following the prescription of Rawlings et al. (2001), i.e. using the
detection of any line from at least a doubly-ionised ion and G=galaxy for which we have little or no spectroscopic
information to classify it as HEG or LEG but one which is spatially
resolved in our imaging data.  {\bf Column 11:} The reference to the
redshift of the source, Jea - Jarvis et al. (2001); Rea - Rawlings et
al. (1996); and deR - de Ruiter et al. (1986). }}
\end{table*}

\section{Notes on Individual Sources}\label{sec:notes}
\makebox{}

\begin{figure*}[!ht]
{\hbox to \textwidth{\epsfxsize=0.45\textwidth
\epsfbox{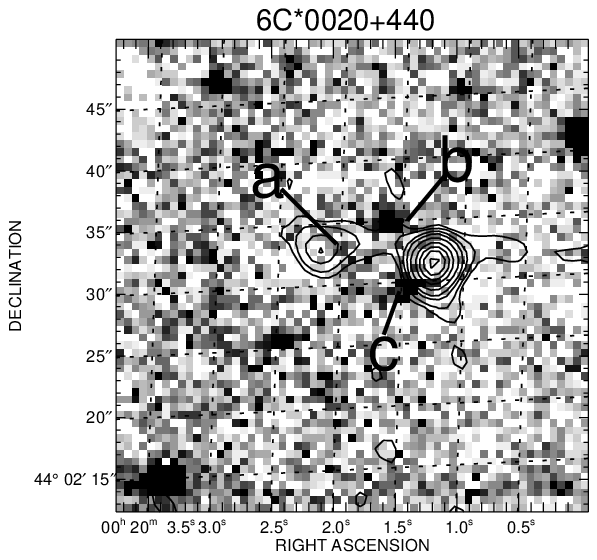}
\epsfxsize=0.45\textwidth
\epsfbox{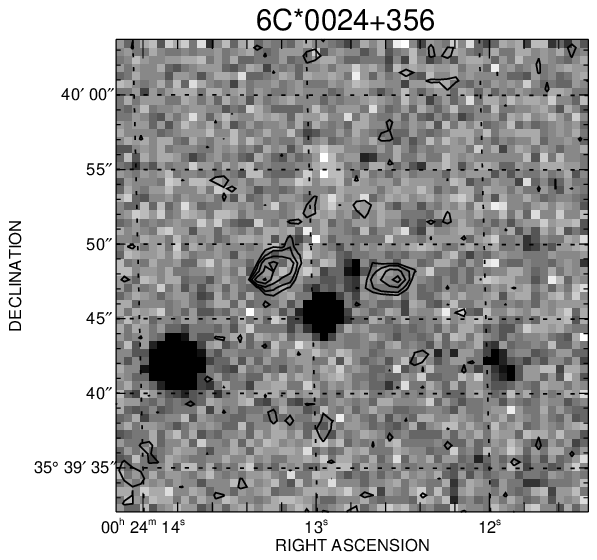} }} {\hbox
to \textwidth{\epsfxsize=0.45\textwidth
\epsfbox{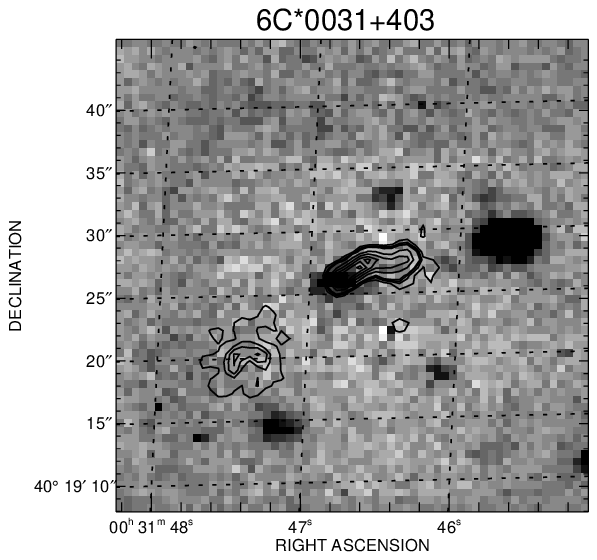}
\epsfxsize=0.45\textwidth
\epsfbox{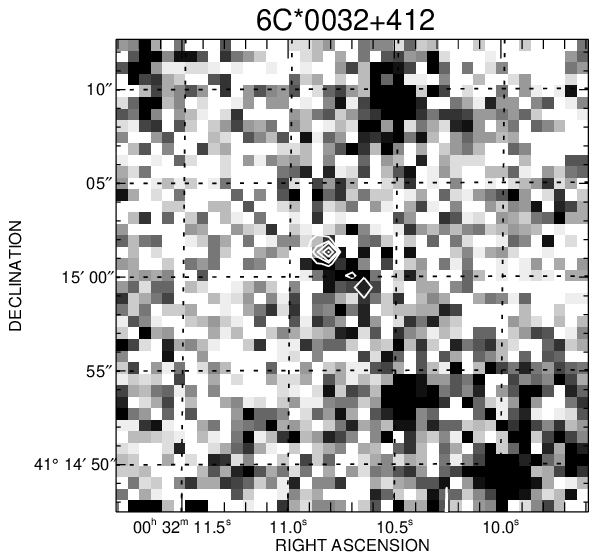} }}
{\caption{\label{fig:kband_images} \kband images (greyscale) of 28 6C*
sources, with the radio contours overlaid. A detailed account of the
radio galaxy 6C*0140+326 can be found in Rawlings et al. (1996) and a
deep near-infrared image in van Breugel et al. (1998). The coordinates are
in B1950 and the frequency of the radio contours are those presentated
in Blundell et al. (1998) except where otherwise stated.  For presentation purposes
the images obtained in 1997 have all been binned up by averaging over
$2 \times 2$ pixels to produce roughly the same pixel scale as the
1993 and 1994 exposures. Likewise the images taken in 2000 have been
binned up by averaging over $3 \times 3$ pixels, this produces a
slightly smaller pixel scale than for the other observations. For the
images of 6C*0050+419, 6C*0106+397 and 6C*0132+330 these have also
been smoothed by a Gaussian with $\sigma = 1$ pixel to make the ID
clearer. For the images where the radio contours are not visible due
to the scale of the optical ID encompassing most of the radio contours
we have marked the peak of the radio emission with a white
cross. There was a flat-fielding problem for the image of 6C*0154+450,
which is apparent towards the north-east of the image but this does
not affect the magnitude given in Table~\ref{tab:kmags}. The \kband
images and overlays of the sources now excluded from 6C* are presented in
Appendix B.}}
\end{figure*}

{\bf 6C*0020+440} ($z = 2.988$) The extremely faint plausible ID, (a) in
Fig.~\ref{fig:kband_images}, is situated between the two radio lobes but
closer to the eastern lobe. There are also two bright objects close to
the radio structure. Both object (b) $\approx 2.5$ arcsec to the north
of our ID, and object (c) $\approx 3$ arcsec to the south-west of this
identification may contribute some flux to the \kband magnitude in
apertures $> 3$\,arcsec.

{\bf 6C*0024+356} ($z = 2.161$)
The ID of this source lies between the two radio lobes but shifted towards
the western lobe. The ID is close to a brighter object making any
magnitude estimation difficult. 

{\bf 6C*0031+403} ($z = 1.619$) Our \kband image shows a bright ID
co-spatial with the flat-spectrum core. The radio structure of this
source is somewhat unusual for the 6C* sample: a one-sided jet emerges
from the bright core, suggestive of a jet disrupted on one side of the
source. We speculate that this object might be a reddened quasar with
an additional non-stellar component to the \kband light, due to the
unusually bright \kband magnitude for a source at this redshift, although
there is no evidence of any broad line component from our spectroscopic observations.

{\bf 6C*0032+412} ($z = 3.658$) Our infrared spectroscopic
observations of this source (Jarvis et al. 2001) reveals that $\sim
50$\% of the \kband flux is contributed by the [OIII] emission lines
at 4959 \AA\, and 5007 \AA\,. The errors on this value are large
because of the tentative nature of the UKIRT ID, and consequently
inaccurate photometry. There are hints that the \kband light is
extended along the radio axis, probably due to significant
contamination by emission lines. The astrometry for this image is
accurate to $\sim 1.5$~arcsec and uncertain enough to reconcile the
apparent displacement of the \kband ID with the radio emission.

{\bf 6C*0041+469} ($z = 2.14$?)  Our $I-$band (see appendix A) and
\kband images reveal highly aligned emission, although the astrometry
for the $I-$band image is offset from the expected position by $\sim
1$\,arcsec to the west with our astrometric fit to the image. The tentative redshift of this object is $z =
2.14$; at this redshift the H$\alpha$ emission line is redshifted into
the \kband window and thus there may be a non-stellar component
contributing to the \kband flux, however, the \kband magnitude is
consistent with the quoted redshift.

{\bf 6C*0050+419} ($z = 1.748$?)  There is a faint object to the south
of the peak of the compact radio emission (a), which is most plausibly
the ID. Deep \iband imaging with the Keck telescope (appendix A) also
reveals a faint blob of
emission to the south of the peak of the radio emission. There also
appears to be a faint blob of emission slightly to the north of the
radio peak in the \iband image, although there is no visible \kband
counterpart we cannot rule this out as being associated with the radio
source.
There is also a faint blob to the south-west (object `b') in both our
$K$- and $I$-band images co-spatial with the faint radio component in
our 8.4 GHz map, which could also be the ID, and has a \kband
magnitude of $K \simeq 19.5$ (8 arcsec aperture).

{\bf 6C*0052+471} ($z = 1.935$)
We find bright, unresolved \kband emission associated with the centre of the radio
emission; the object is a quasar.

{\bf 6C*0058+495} ($z = 1.173$)
A bright \kband ID is co-spatial (within the astrometric uncertainty)
with the centre of the radio lobes,
although no core component is present in our radio maps.

{\bf 6C*0106+397} ($z = 2.284$) We find faint \kband emission situated
between the two radio lobes, and we take this as the ID. There is much
brighter emission to the south-east of the southern most radio
lobe with $K = 17.8$ (8 arcsec aperture) which we associate with a
$z = 0.632$ foreground galaxy (Jarvis et al. 2001). 

{\bf 6C*0112+372} ($z = 2.535$)
This source is spatially resolved in 0.9 arcsec seeing at the position
of the bright radio emission, and is therefore identified as a radio
galaxy. The peak of the radio emission is represented by the white
cross in Fig.~\ref{fig:kband_images}.

{\bf 6C*0115+394} ($z = 2.241$)
Our \kband image shows a resolved identification in 0.8 arcsec seeing, co-spatial with the radio core.

{\bf 6C*0118+386} ($z = 2.350$) The identification of this source is
coincident with the central radio component which has a spectral index
of $\alpha \sim 0.5$, and is taken to be the site of the AGN. The
\kband magnitude of this object is not unreasonable for the redshift,
but we cannot eliminate the possibility that it is contaminated by a
foreground $z = 0.529$ galaxy seen in emission lines in the
spectrum. This galaxy is sufficiently far from either of the radio
lobes that significant gravitational lensing of the flux-density appears unlikely. We note three bright
resolved galaxies near the northern hotspot which may be part of an
intervening $z = 0.529$ group or cluster.

{\bf 6C*0122+426} ($z = 2.635$)
Our \kband image shows faint emission associated with the radio
source. Our astrometry is not good enough to determine whether the ID
is on either of the bright radio components or between them.

{\bf 6C*0128+394} ($z = 0.929$)
Our \kband image shows a bright ID coincident with the centre of the double lobed radio emission.

{\bf 6C*0132+330} ($z = 1.710$)
We find a very faint \kband ID associated with the centroid of the radio emission.

{\bf 6C*0133+486} ($z = 1.029$?)  There are two possible
identifications for this radio source, although these are along the
line of the radio jet and both may be associated with the radio
galaxy. Our tentative redshift of $z = 1.029$ from the brightest
component (a) leads us to conclude, from the $K-z$ diagram, that this
is the host galaxy, although this is still $\sim 0.8$ magnitudes
fainter than the best-fit line through the 6CE/6C* $K-z$ diagram
(Fig.~\ref{fig:kzdiag}). However, object (b) $\sim 2$ arcsec to the
north cannot be ruled out as the host and further spectroscopic
observations will be needed to confirm the ID and redshift of this
radio source.

{\bf 6C*0135+313} ($z = 2.199$)
The \kband ID of this source lies at the centre of the bright radio
emission. There is also a brighter source ($K = 18.9$ in an 8~arcsec
aperture) to the south-east which is possibly a foreground object,
although we have no spectroscopic data to confirm this.

{\bf 6C*0136+388} ($z = 1.108$?)
The \kband ID of this source lies at the radio core. The radio core
is not visible in the image presented in Fig.~\ref{fig:kband_images}
because of the bright \kband ID; the peak in the radio emission
is represented by the white cross. The \kband magnitude is also
consistent with the $K-z$ relation at this redshift.

{\bf 6C*0139+344} ($z = 1.637$) Our optical spectra reveal a nearby
foreground galaxy at $z = 0.37$ which could be contributing to our
measured \kband magnitude. Our \kband image reveals two galaxies
towards what appears to be a steep-spectrum radio lobe to the
north. We take the fainter of these (a) to be the radio galaxy with the $z
= 0.37$ foreground galaxy (b) approximately 2.5 arcsec to the north, which
is in agreement with our astrometry from the 2D spectrum.  There are
no objects in our images which are coincident with the faint radio
component in our 4.9 GHz map at 01 39 25.87 +34 27 02.5 (B1950) which
we originally thought to be the core. 

{\bf 6C*0140+326} ($z = 4.41$)
This gravitationally lensed source is discussed in detail in Rawlings et al. (1996).

{\bf 6C*0142+427} ($z = 2.225$)
The \kband ID for this source is coincident with the radio core.

{\bf 6C*0152+463} ($z = 2.279$)
The \kband ID of this source lies between the two radio lobes.

{\bf 6C*0154+450} ($z = 1.295$) We find an unresolved point source co-spatial
with the radio core, as expected for a quasar.

{\bf 6C*0155+424} ($z = 0.513$) The $R-$ (Jarvis 2000) and \kband
images of this radio source reveal three optically bright sources that
could each plausibly be the radio galaxy. The \kband magnitudes of all
three point towards a low redshift ($z < 1$) and our optical
spectroscopy revealed that the ID closest to the centre of the radio
emission has a redshift $z = 0.513$ and is probably the host galaxy on
positional grounds. The nearest neighbour to this source is $\ltsim 1.5$\,arcsec
to the north-west, and the similar \kband magnitudes suggest that
these two objects could be in the process of merging. There is also
another optical source $\approx 4$ arcsec to the south-east of these
two objects. This also has a similarly bright \kband magnitude and is
also plausibly a true companion galaxy. There is faint diffuse
emission surrounding the two closest objects, and the third has faint
emission reaching towards the other two, which is also indicative that
this system may be undergoing a major merger. Further spectroscopy in
good seeing will be required to ascertain whether they are merging,
and deeper imaging may find more diffuse emission surrounding all
three objects within the radio emission.

{\bf 6C*0158+315} ($z = 1.505$)
We find a \kband ID co-spatial with the compact radio source. This
was initially thought to be a hot-spot associated with a larger
source. However, the imaging data presented in this paper suggest that
it is indeed a discrete source, although a triple source spanning
$\sim 14$ arcmin cannot be ruled out completely (but see Jarvis et al. 2001).

{\bf 6C*0201+499} ($z = 1.981$)
The \kband ID for this source is coincident with the southern lobe,
although the small angular size of this source ($\theta \approx 1$
arcsec) means that astrometric uncertainties could easily place the ID
at the centre of the lobes.

{\bf 6C*0202+478} ($z = 1.613$?)
The ID (a) for this source is in close proximity to two bright stars. The
ID is faint at the position of the weak radio emission, between the two
radio lobes, which we now take to include the core.

{\bf 6C*0208+344} ($z = 1.920$)
We find a very faint \kband identification at the position of the southern radio
component, although there are no indications from the radio data that
this has a flat spectrum.

{\bf 6C*0209+276} ($z = 1.141$?)
Our 4.9 GHz radio map barely resolves this source. We do find a
bright \kband identification at the position of the radio source, its
magnitude suggests that our tentative redshift of $z = 1.141$ is correct.

\addtocounter{figure}{-1}

\begin{figure*}
{\hbox to \textwidth{\epsfxsize=0.45\textwidth
\epsfbox{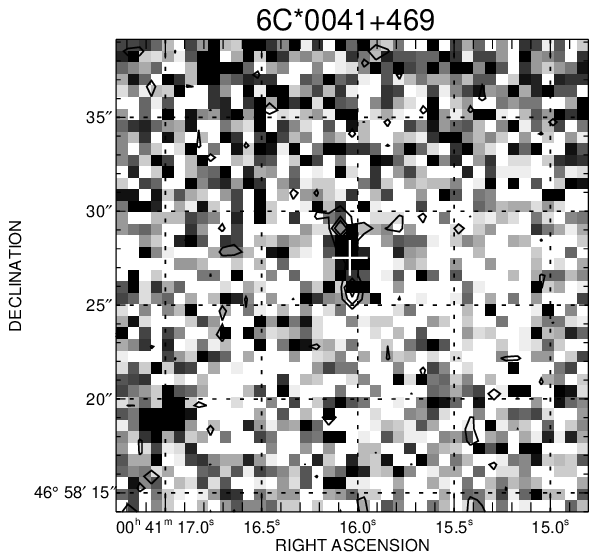}
\epsfxsize=0.45\textwidth
\epsfbox{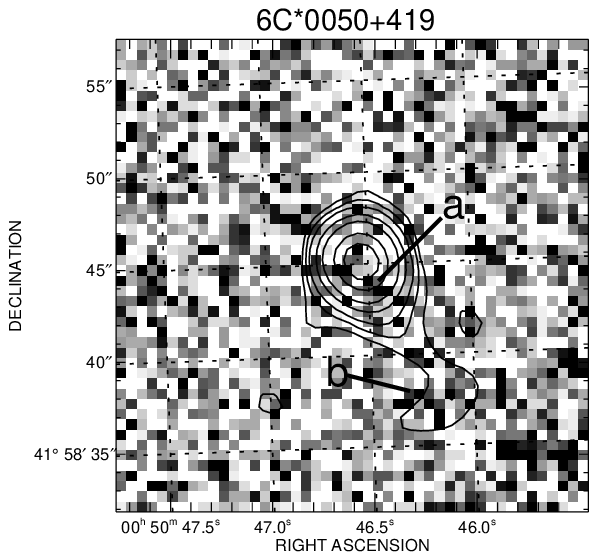} }}
{\hbox to \textwidth{\epsfxsize=0.45\textwidth
\epsfbox{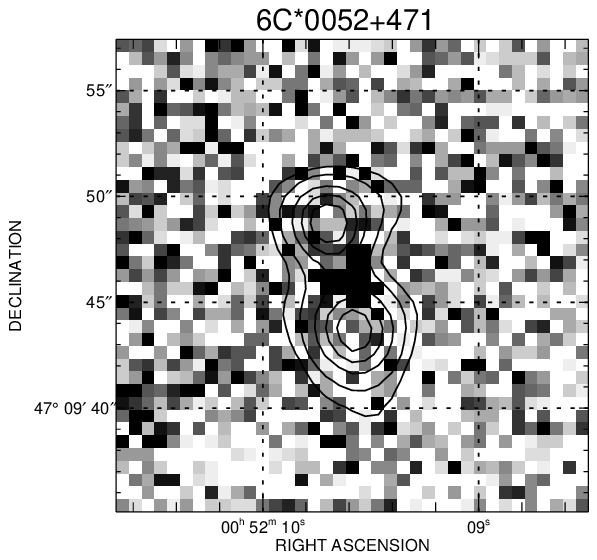} 
\epsfxsize=0.45\textwidth
\epsfbox{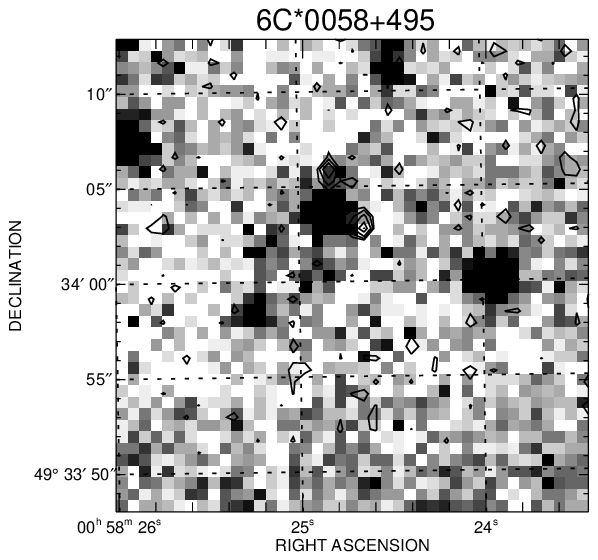} }}
{\hbox to \textwidth{\epsfxsize=0.45\textwidth
\epsfbox{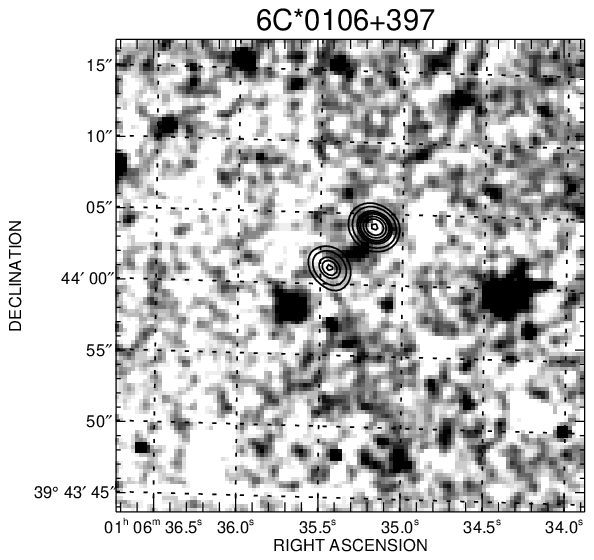} 
\epsfxsize=0.45\textwidth
\epsfbox{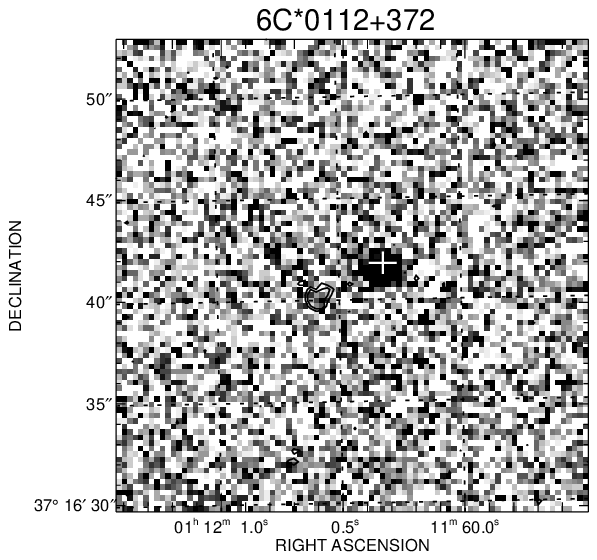} }}
{\caption{\label{fig:ov2} cont.}}
\end{figure*}

\addtocounter{figure}{-1}

\begin{figure*}
{\hbox to \textwidth{\epsfxsize=0.45\textwidth
\epsfbox{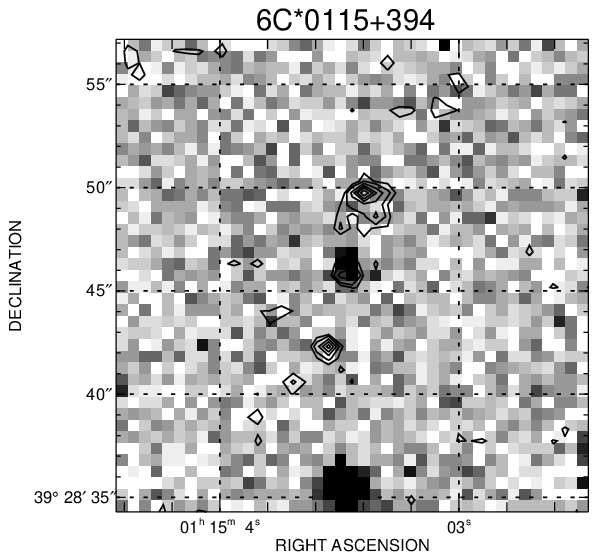}
\epsfxsize=0.45\textwidth
\epsfbox{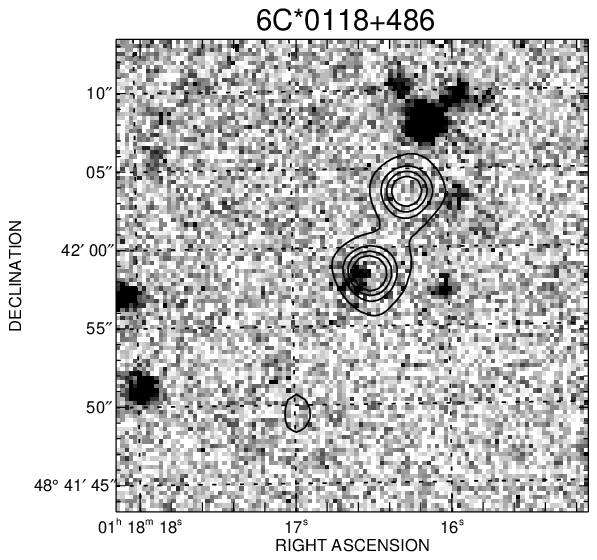} }}
{\hbox to \textwidth{\epsfxsize=0.45\textwidth
\epsfbox{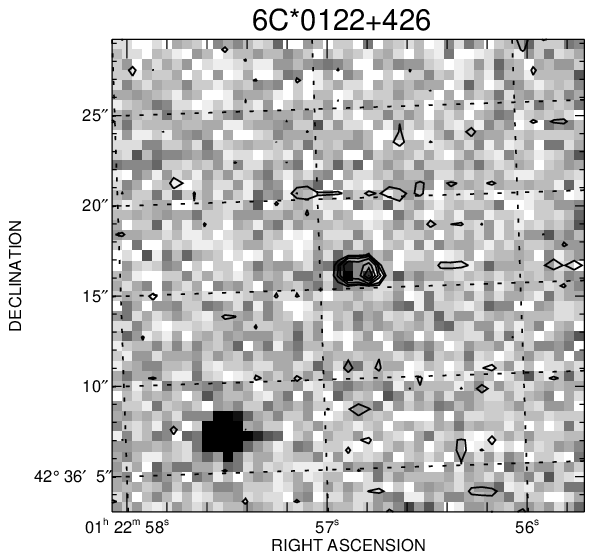}
\epsfxsize=0.45\textwidth
\epsfbox{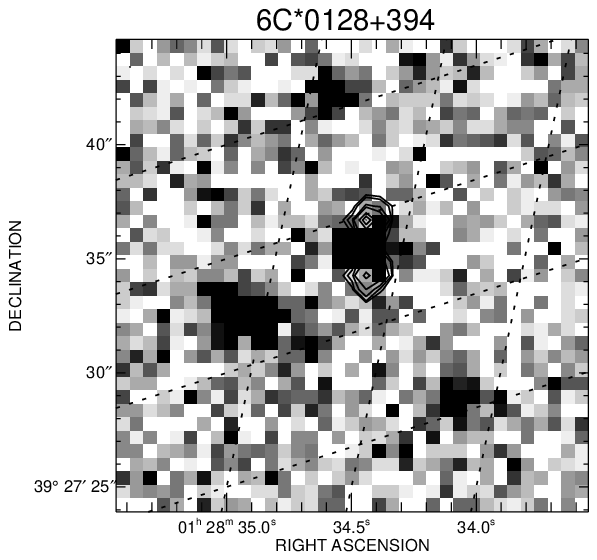} }}
{\hbox to \textwidth{\epsfxsize=0.45\textwidth
\epsfbox{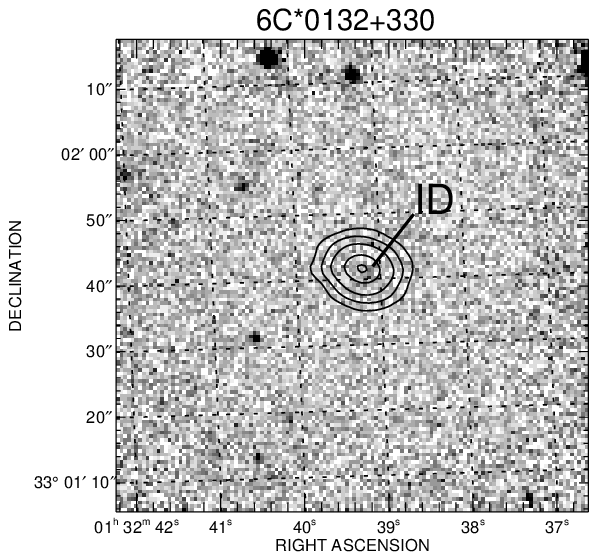}
\epsfxsize=0.45\textwidth
\epsfbox{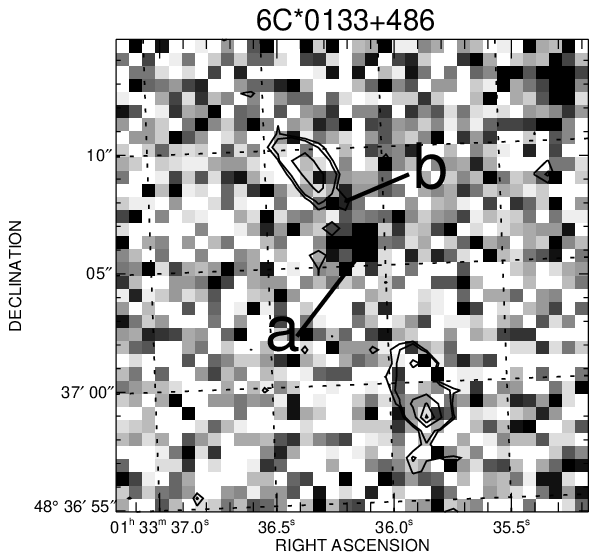} }}
{\caption{\label{fig:ov3} cont.}}
\end{figure*}

\addtocounter{figure}{-1}

\begin{figure*}
{\hbox to \textwidth{\epsfxsize=0.45\textwidth
\epsfbox{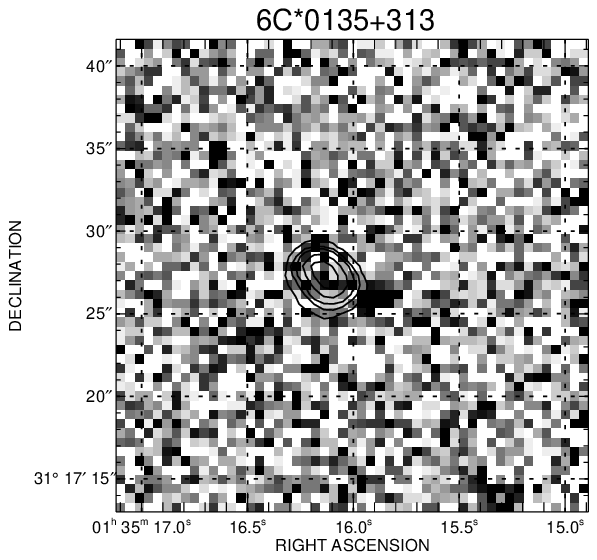} 
\epsfxsize=0.45\textwidth
\epsfbox{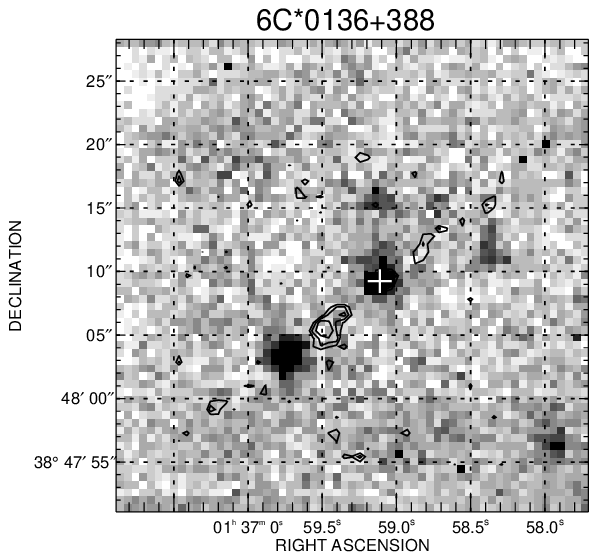} }}
{\hbox to \textwidth{\epsfxsize=0.45\textwidth
\epsfbox{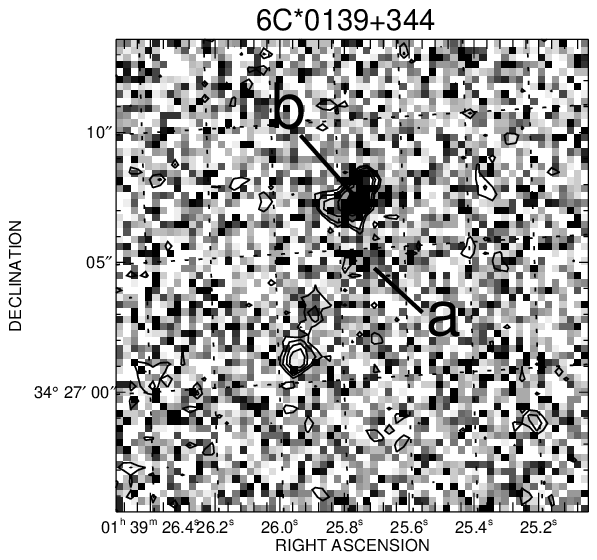}
\epsfxsize=0.45\textwidth
\epsfbox{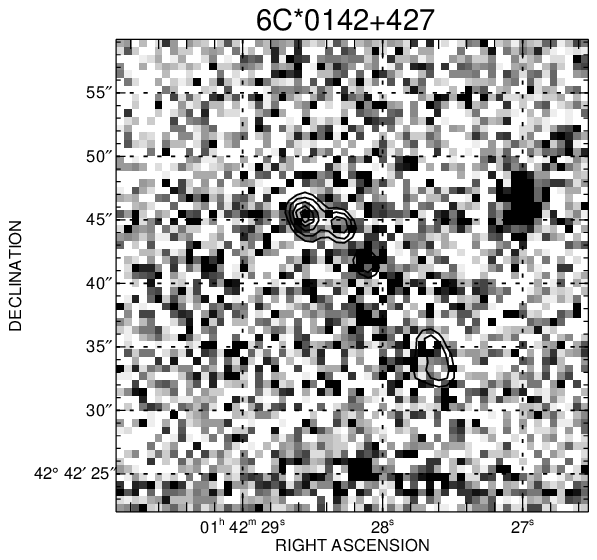} }}
{\hbox to \textwidth{\epsfxsize=0.45\textwidth
\epsfbox{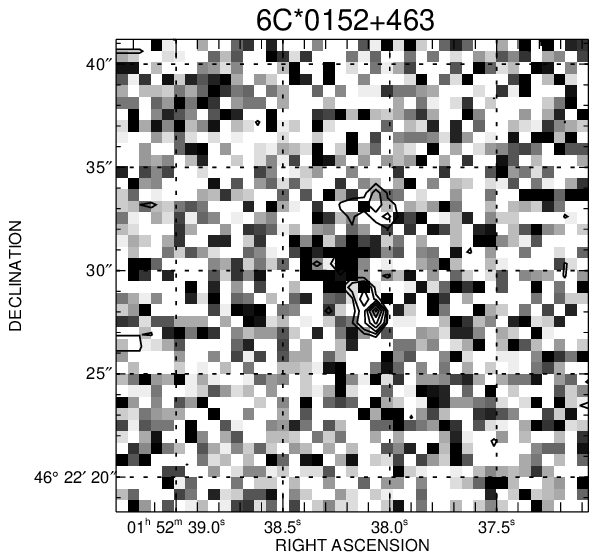}
\epsfxsize=0.45\textwidth
\epsfbox{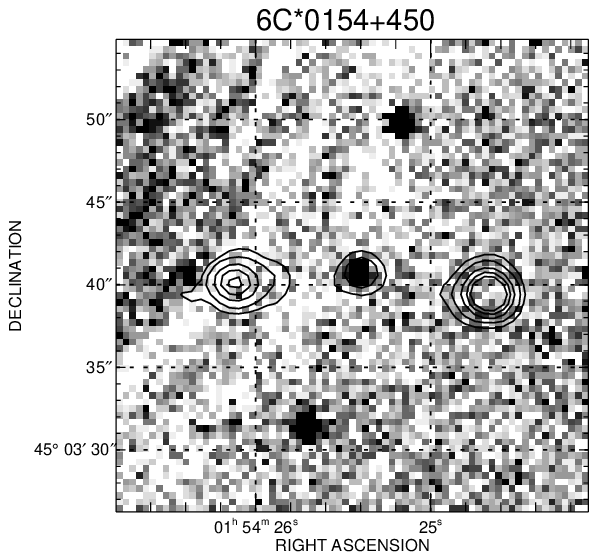} }}
{\caption{\label{fig:ov4} cont.}}
\end{figure*}

\addtocounter{figure}{-1}

\begin{figure*}
{\hbox to \textwidth{\epsfxsize=0.45\textwidth
\epsfbox{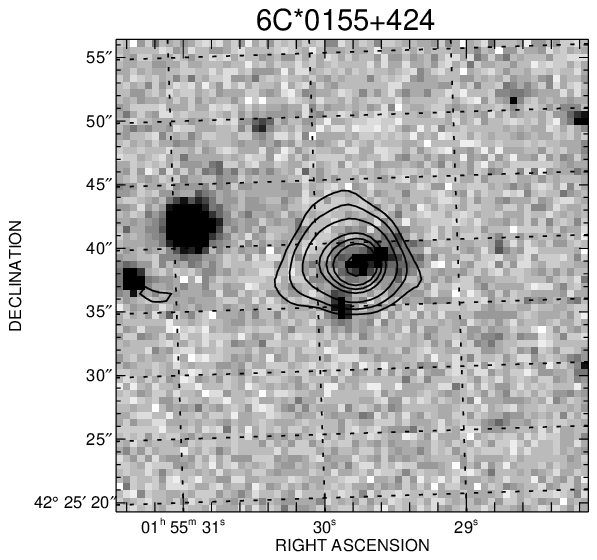} 
\epsfxsize=0.45\textwidth
\epsfbox{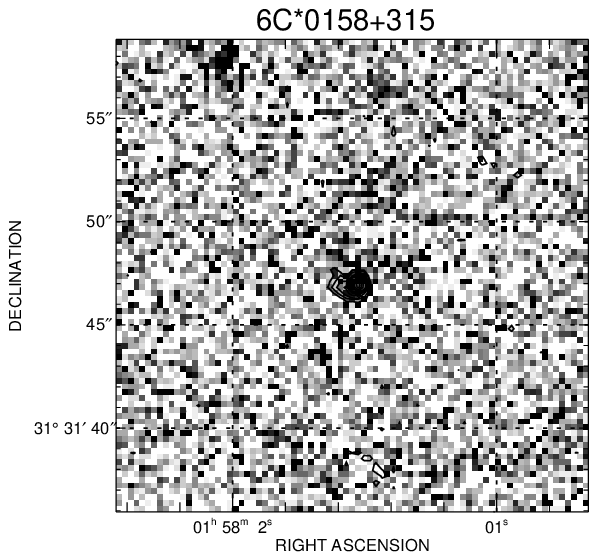} }}
{\hbox to \textwidth{\epsfxsize=0.45\textwidth
\epsfbox{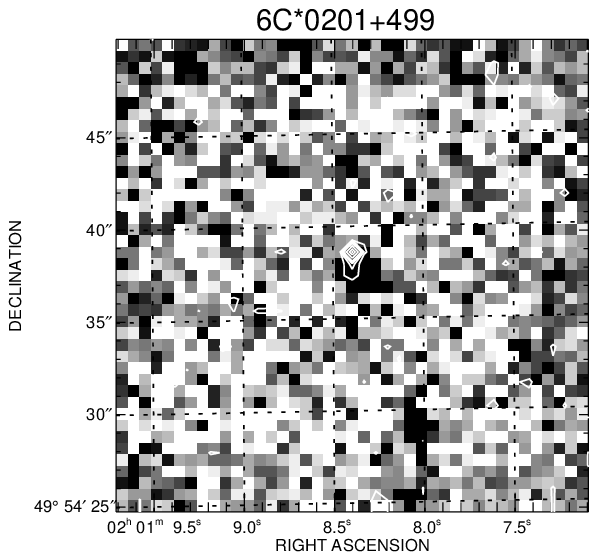}
\epsfxsize=0.45\textwidth
\epsfbox{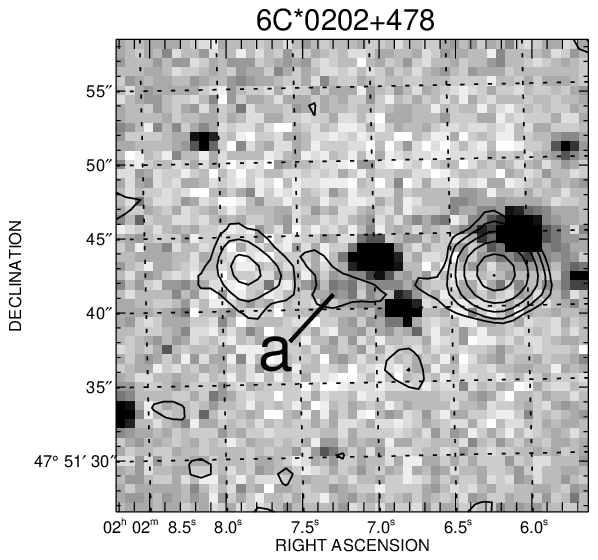} }}
{\hbox to \textwidth{\epsfxsize=0.45\textwidth
\epsfbox{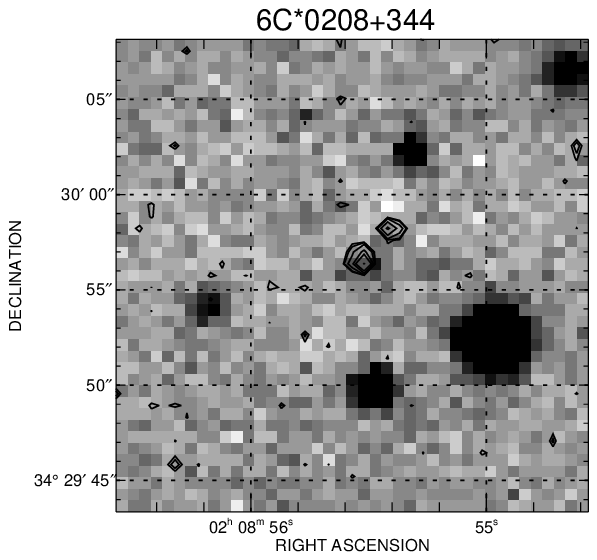}
\epsfxsize=0.45\textwidth
\epsfbox{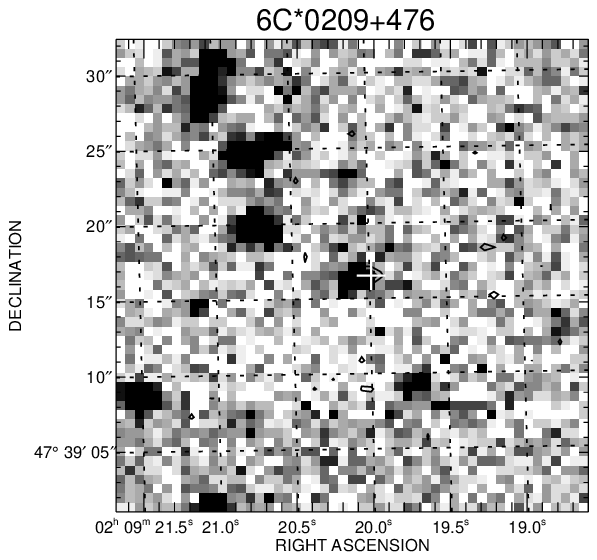} }}
{\caption{\label{fig:ov5} cont.}}
\end{figure*}

\section{The $K-$\lowercase{$z$} Hubble diagram}\label{sec:kz}
In this section we compare the infrared Hubble diagram for the 6C*
sources with previous flux-density-limited samples, namely the 3CRR
sample and 6CE sample along with data from the 7C-III sample (Lacy et
al. 2000). The 3CRR (6CE) sample has a flux-density limit a factor of
about twelve (two) brighter than 6C*, this range allows us to break
the strong biases inherent in flux-density limited samples
(e.g. Blundell et al. 1999). The $K$-band magnitudes for the 3CRR
sample have been compiled from the literature, the majority from the
work of Lilly \& Longair (1984). Other sources of 3CRR $K$-band
magnitudes are Best, Longair \& R\"ottgering (1998) and de Vries et
al. (1998). These data are complemented by the virtually
spectroscopically complete 6CE sample. The $K$-band magnitudes for
this sample have been gathered from the literature, mainly from Eales et al. (1997)
and Eales \& Rawlings (1996). There are five sources from the 6C*
sample which are omitted from the $K-z$ analysis: these are the two
broad line quasars (6C*0052+471 and 6C*0154+450) where the \kband
light may be dominated by emission from the quasar nucleus,
6C*0139+344 which may have a foreground galaxy contributing to its
\kband flux, 6C*0155+424 as this looks like a merger
system and the radio source ID is uncertain and 6C*0202+478 which is
in too close proximity to a star to measure accurate photometry. This
source is also one of the sources with an uncertain redshift, thus
there are only five filled stars in Figs.~\ref{fig:kzdiag} \&~\ref{fig:Kzmod}.

\subsection{Correcting to a metric aperture}
A problem when comparing broad-band magnitudes for objects over a
large redshift baseline is that the magnitudes canot be measured in
an aperture of a given physical diameter, if the redshift is
unknown. If the redshift is known then the angular diameter can then
be adjusted to correspond to a metric aperture of a given physical
size for all of the sources. However, we do not have the \kband images
for the sources in the 3CRR and
6CE samples
available for re-measuring the photometry in a single metric aperture.
Therefore, we need a method of converting from our broad-band
magnitudes measured in a specific angular aperture to our chosen metric
aperture. To correct to a metric aperture, we follow the prescription
of Eales et al. (1997), where the metric aperture used is 63.9 kpc, which
corresponds to an angular size of $\approx 8$ arcsec at $z \geq 1$. For the low-redshift ($z <
0.6$) sources we use the curve of growth analysis of Sandage (1972),
because at these redshifts the radio galaxies are predominantly
associated with giant ellipticals (e.g. Rogstad \& Ekers 1969). At
high redshift ($z > 0.6$) we use
a power-law intensity profile of the form $I \propto r^{\alpha}$ where
we use $\alpha = 0.35$, the value derived by Eales et al. (1997).

\subsection{Emission-line contamination of $K$-band magnitudes}\label{sec:emline}
The emission-line contamination of the measured continuum magnitudes
for our dataset also needs to be accounted for, especially for the
most-luminous sources at high redshift, where the bright optical
emission lines are redshifted into the infrared. 

\begin{figure}[!ht]
{\hbox to 0.5\textwidth{\epsfxsize=0.45\textwidth
\epsfbox{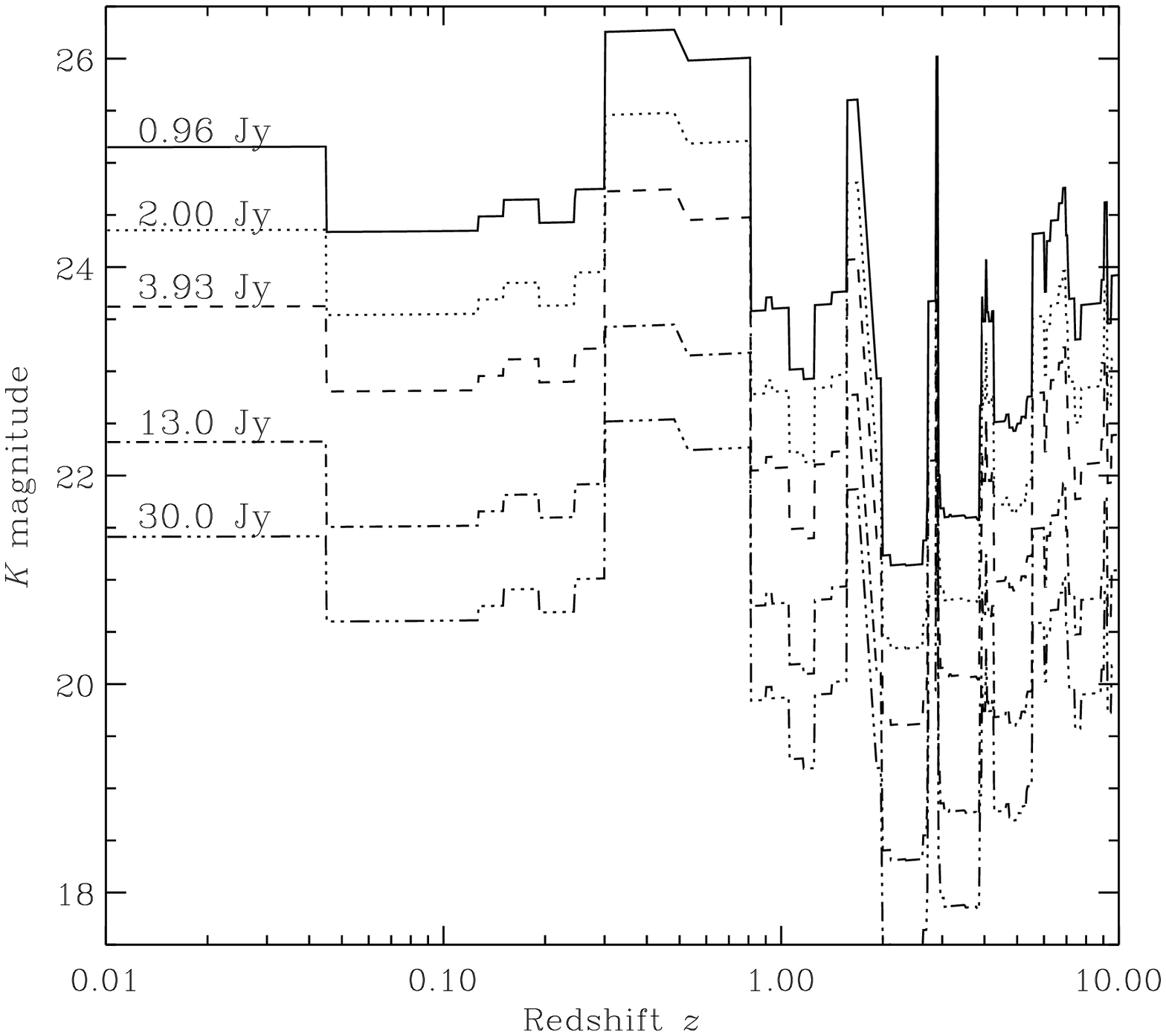} }}
{\caption{\label{fig:line_cont} Emission line contribution to the
\kband magnitudes for various radio flux-densities assuming the
power-law relation of $L_{\rm [OII]} \propto L_{151}^{1.00}$.  The
troughs in the plot occur where bright emission lines are redshifted
into the \kband transmission window.  For the samples used in this
paper, the 3CRR sample is the dominant contributor of sources at $z <
1.8$ and thus the 13~Jy line through the plot is an estimate of the
minimum contribution from emission lines to the \kband flux. At $z >
2$ the 6CE and 6C* samples dominate and as such the upper three lines
from 3.93~Jy to 0.96~Jy demonstrate the typical contribution from
emission lines at these redshifts. }}
\end{figure}

To subtract this contribution we use the correlation between [OII]
emission-line luminosity $L_{\rm [OII]}$ and the low-frequency radio
luminosity $L_{151}$ from Willott (2000), where $L_{\rm [OII]} \propto
L_{151}^{1.00 \pm 0.04}$. Thus, for a source with a given $L_{151}$ we
can estimate $L_{\rm [OII]}$. Then by using the emission-line flux
ratios for radio galaxies from McCarthy (1993) for emission lines
below 5007\AA\,, and from Robinson et al. (1987), Ward et al. (1991)
and Rudy et al. (1992) for the longer wavelength emission lines, we
are able to determine the contribution to the \kband magnitude from all of the
other emission-lines. Assuming a square transmission window centred
at 2.2$\mu$m with a width of 0.48$\mu$m, and a flux-density of $4.07
\times 10^{-14}$ W m$^{-2}$\,\AA$^{-1}$\, corresponding to a zeroth
magnitude star in the \kband (Longair 1992), we convert this
flux-density to a $K$-magnitude. In Fig.~\ref{fig:line_cont} we show
the emission-line contamination to the \kband magnitude for various
radio flux-densities over the redshift range $0.0 < z < 10$. The
total emission-line contribution to the \kband flux is subtracted from
the measured \kband flux and the remaining flux should be due to the
stellar continuum. This assumes that there is no other contribution from the
non-stellar continuum such as reddened and/or scattered quasar light
(e.g. Simpson, Rawlings \& Lacy 1999; Leyshon \& Eales 1998). 

\begin{figure*}[!ht]
{\hbox to 1.\textwidth{\epsfxsize=1.\textwidth
\epsfbox{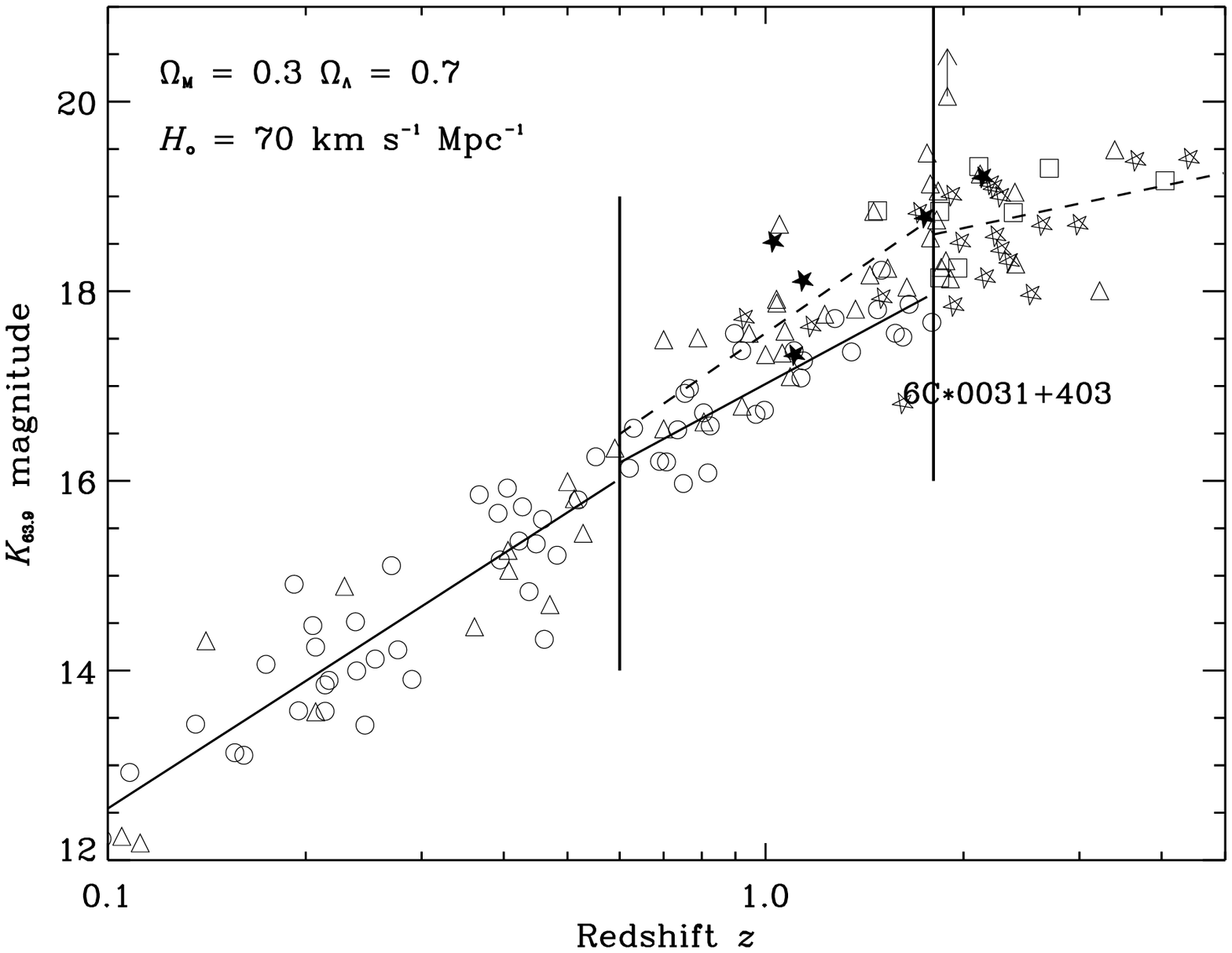}}}
{\caption{\label{fig:kzdiag} The $K-z$ Hubble diagram for radio
galaxies from the 3CRR (circles), 6CE (triangles), 6C* (stars) and
7C-III (squares) samples, for a $\Omega_{\rm M} = 0.3$,
$\Omega_{\Lambda}=0.7$ cosmology with $H_{\circ} = 70\,{\rm
km\,s}^{-1}\,{\rm Mpc}^{-1}$, in line with the recent observational
evidence (e.g. Perlmutter et al. 1999; Balbi et al. 2000; Gibson \&
Stetson 2001).  $K_{63.9}$ denotes the $K-$band magnitude within a
metric aperture of 63.9 kpc (c.f. Eales et al. 1997), and all
magnitudes have been corrected for emission-lines following the
prescription given in Sec.~\ref{sec:emline}.  The two vertical lines
show the redshift above which the alignment effect begins to be seen
($z = 0.6$) and the higher redshift at which we chose to split the
data and beyond which there are zero 3CRR sources ($z = 1.8$). The
solid lines are the fits to the 3CRR data points at $z < 0.6$ and $0.6
< z < 1.8$. The dashed line is the fit to the 6CE and 6C* sources at
$0.6 < z < 1.8$ and $z > 1.8$. 6C*0031+403 probably has an AGN
component contributing to the $K-$band magnitude
(Sec.~\ref{sec:notes}) and is labelled. The filled stars represent
five of the six objects in 6C* which do not yet have completely secure redshifts.}}
\end{figure*}

Fig.~\ref{fig:kzdiag} shows the emission-line corrected $K-z$ digram
for the objects in our dataset. To analyse the differences between
samples at low and high redshifts we split the data into three
redshift bins. The low-redshift bin ($z < 0.6$) is the low-redshift
bin chosen by Eales et al. (1997), and is the redshift above which the alignment
effect is seen (McCarthy 1993). The medium-redshift bin ($0.6 < z
<1.8$) was chosen to enable us to directly compare the 3CRR and 6CE
objects which have the same redshift but vastly different radio
luminosities. This bin also contains nine objects from 6C* at the lower
radio flux-density of $0.96 \leq S_{151} \leq 2.00$ Jy. The
high-redshift bin ($z > 1.8$) contains ten 6CE source and fifteen 6C*
objects. We also add seven objects from the
7C-III sample from Lacy et al. (2000) to the high-redshift bin.

\section{Galaxy evolution and the formation epoch}\label{sec:mainevolution}

\subsection{Dispersion in the $K-z$ diagram}\label{sec:dispersion}

We now calculate the residuals in \kband magnitude from straight line
fits (calculated by minimising the square of the residuals in
magnitude) to various subsets of the data in the three redshift bins
considered by Eales et al. (1997), $z < 0.6$, $0.6 < z < 1.8$ and $z > 1.8$. To
avoid the possibility of confusing an increase in the dispersion in
the data with the correlation between $L_{151}$ and the \kband
magnitude (c.f. Dunlop \& Peacock 1993; Eales et al. 1997) we fit separate
straight lines to the 3CRR data points and to the 6CE and 6C* data
points which we consider together due to their similar flux-density
limits. For the following sections concerning the $K-z$ diagram we
also switch to the currently favoured cosmology of \costwo ~with
$H_{\circ} = 70\,{\rm km\,s}^{-1}\,{\rm Mpc}^{-1}$. Although this
choice does not affect the dispersion calculated in this section (as
all of the magnitudes have been adjusted to the chosen matric aperture
in this cosmology), it
does affect the results presented in Sec.~\ref{sec:evolution} where
synthetic spectral libraries are used to generate the expected path
through the $K-z$ diagram for various star-formation
histories. 

For the low-redshift sources in the 3CRR sample, we find that the
dispersion about the fitted line has a standard deviation of $\sigma =
0.52$. Repeating this for the 6CE sources at $z < 0.6$ we find that
the standard deviation is slightly higher with $\sigma = 0.57$. The
$F$-variance test for these distributions shows that they are
statistically indistinguishable, with a significance of $\sim 72$\%
that they are drawn from different distributions, agreeing with the
results of Eales et al. (1997).

At higher redshift we find that the standard deviation of the
residuals in \kband magnitude for 3CRR sources at medium redshift
($0.6 < z < 1.8$) is $\sigma = 0.36$. The $F$-variance test shows that
the difference in dispersion between the low- and medium-redshift 3CRR
bins are significantly different at the 97\% level. Hence, we find
less scatter between $0.6 < z < 1.8$ than at $z < 0.6$ for the 3CRR
sources in our sample. This can be explained in two ways. First, if
there is a radio luminosity dependent component other than
emission-lines, contributing to the \kband flux, then the lower
luminosity host galaxies will be more affected than the hosts with
higher luminosities galaxies. Consequently the spread in observed magnitude becomes
tighter, however this contribution would have to be significant ($> 30\%$) and
the observational evidence suggests that this is not the case at least
at $z \sim 1$ (e.g. Leyshon \& Eales 1998; Simpson et al. 1999).  Second, we know that the 3CRR galaxies in this redshift bin are
the most extreme radio luminous sources in the universe. Therefore, we
would expect them to harbour the most massive black holes, formation
of which probably requires very tight physical conditions
intrinsically linked to the host galaxy. Thus, the tightness of the
$K-z$ relation for 3CRR sources in the mid-redshift bin may be providing an
insight into the physical conditions needed to form the most massive
black holes and consequently the most luminous radio galaxies,
whereas lower-luminosity radio galaxies with lower-mass black holes
could possibly form in galaxies with a range of properties and
environments, thus the dispersion is higher.

For the medium-redshift 6CE and 6C* points we find that the standard
deviation of the 6CE and 6C* sources about the fitted line has a
value of $\sigma = 0.59$ ($\sigma = 0.55$ if the sources with
uncertain redshifts are omitted from the analysis), compared to $\sigma =
0.57$ for the low-redshift 6CE sources and $\sigma = 0.36$ for the
medium-redshift 3CRR sources. Therefore the dispersion in the
medium-redshift 6CE sources is similar to that at low-redshifts,
whereas there is significant difference (confidence $> 97$\%) between the 3CRR and 6CE/6C*
sources at medium redshifts. This may be explained by both
of the points made earlier to the lower-luminosity 6CE/6C*
sources. The range in host galaxy properties could now span a very
wide range while still harbouring a massive black hole and the radio
luminosity is so much lower that any near-infrared component linked to the
luminosity of the radio emission will be much weaker. 

We now repeat this analysis for the high-redshift ($z > 1.8$)
points. As there are no 3CRR points above $z=1.8$ we only compare the
residuals of the 6CE and 6C* points in the medium- and high-redshift
bins. The standard deviation in the high-redshift bin is $\sigma =
0.51$ (exclusion of the sources with uncertain redshifts does not
alter this value) compared to $\sigma = 0.59$ in the medium-redshift bin. These
are statistically indistinguishable with the $F$-variance test showing
that the variances are different at the 40\% significance level. It is
also worth noting that there is no significant difference in the
normalisation of the fitted line or in the dispersion of the 6CE sources
and 6C* sources at any redshift. Thus, there is no detectable
difference between the two samples.

Therefore, unlike the results of Eales et al. (1997) we find no evidence of an
increase in the dispersion in the \kband magnitudes from $z < 2$ to $z
> 2$. Lacy et al. (2000) also find that the dispersion in luminosities
of the hosts radio galaxies increases towards high redshift. They find
a standard deviation of $\sigma = 0.53$ in the medium-redshift
bin. However, in their high-redshift bin they find that the standard
deviation increases to $\sigma = 0.93$. This difference cannot be
explained by accounting for the emission-line contribution as Lacy et
al. also compensated for this.  One possible effect that we have
neglected is any non-stellar emission, which is not in the form of
emission lines, which was addressed by Lacy et al. (2000).
However, any additional luminosity dependent contribution would also
affect the results of Eales et al. (1997) in the same way, and it
would need to be considerable to make the dispersion of the medium-
and high-redshift distributions significantly different. Thus the lack
of sources at $z > 1.8$ in the work of Eales et al. (1997) and Lacy et
al. (2000) may be playing the dominant r\^ole in their determination of the
dispersion at high redshift.  Therefore, we conclude that there is no
evidence from our data of an increase in the scatter in the $K - z$
diagram from $z \sim 0.6$ to $z \sim 3$ for the 6CE/6C* sources.

\subsection{The $K-z$ digram and evolutionary models}\label{sec:evolution}

In this section we compare our $K-z$ diagram for radio sources from
the 3CRR, 6CE, 7C-III and 6C* radio samples with previous results in
the literature and the galaxy evolution models of Bruzual \& Charlot
(1993).  

Eales et al. (1997) found that in an $\Omega_{\rm M} = 1$ universe
with $\Omega_{\Lambda} = 0$, the hosts of radio galaxies in the 6CE
sample were consistent with a non-evolving population for $z <
2$. However, for a low-density universe ($\Omega_{\rm M} = 0$) the
$K-z$ diagram for the 6CE sample becomes consistent with a passively
evolving population to $z \sim 2$.  Beyond $z > 2$ the radio galaxies
are systematically brighter than the no-evolution model and the
passive evolution curves in both cosmologies considered. Eales et
al. (1997) used this as evidence that we are probing the epoch of
galaxy formation at $z \gtsim 2$. Using high-redshift radio galaxies
from a number of flux-density-limited radio samples, van Breugel et
al. (1998) find that the near-infrared colours of radio galaxies at $z
> 3$ are very blue, consistent with a young stellar population. They
also suggest that the size of the near-IR emission regions at $z > 3$
are comparable with the size of the radio structures, with the near-IR
component also more aligned with the radio components at $z> 3$ than
at $z < 3$. This suggests an intrinsic link between the star-formation
activity and the radio emission at high-redshift, with these two
processes probably triggered at roughly the same time by the same
mechanism.  Lacy et al. (2000) using the flux-density-limited 7C-III
radio sample find evidence that the hosts of radio-loud galaxies
become more luminous with redshift, consistent with a passively
evolving population which formed at high-redshift ($z > 3$). Thus, all
of the evidence suggests that radio
galaxies at $z\, \ltsim\, 3$ are associated with host galaxies which
formed the bulk of their stellar populations at epochs corresponding
to $z\, \gtsim\, 3$ and which have undergone simple passive stellar evolution
from the time the bulk of their stars formed to the time at which they
develop the jets we see (although non-stellar contamination may mean
this inference from \kband magnitudes is not yet firm).

\begin{figure*}[!ht]
{\hbox to 1.\textwidth{\epsfxsize=1.\textwidth
\epsfbox{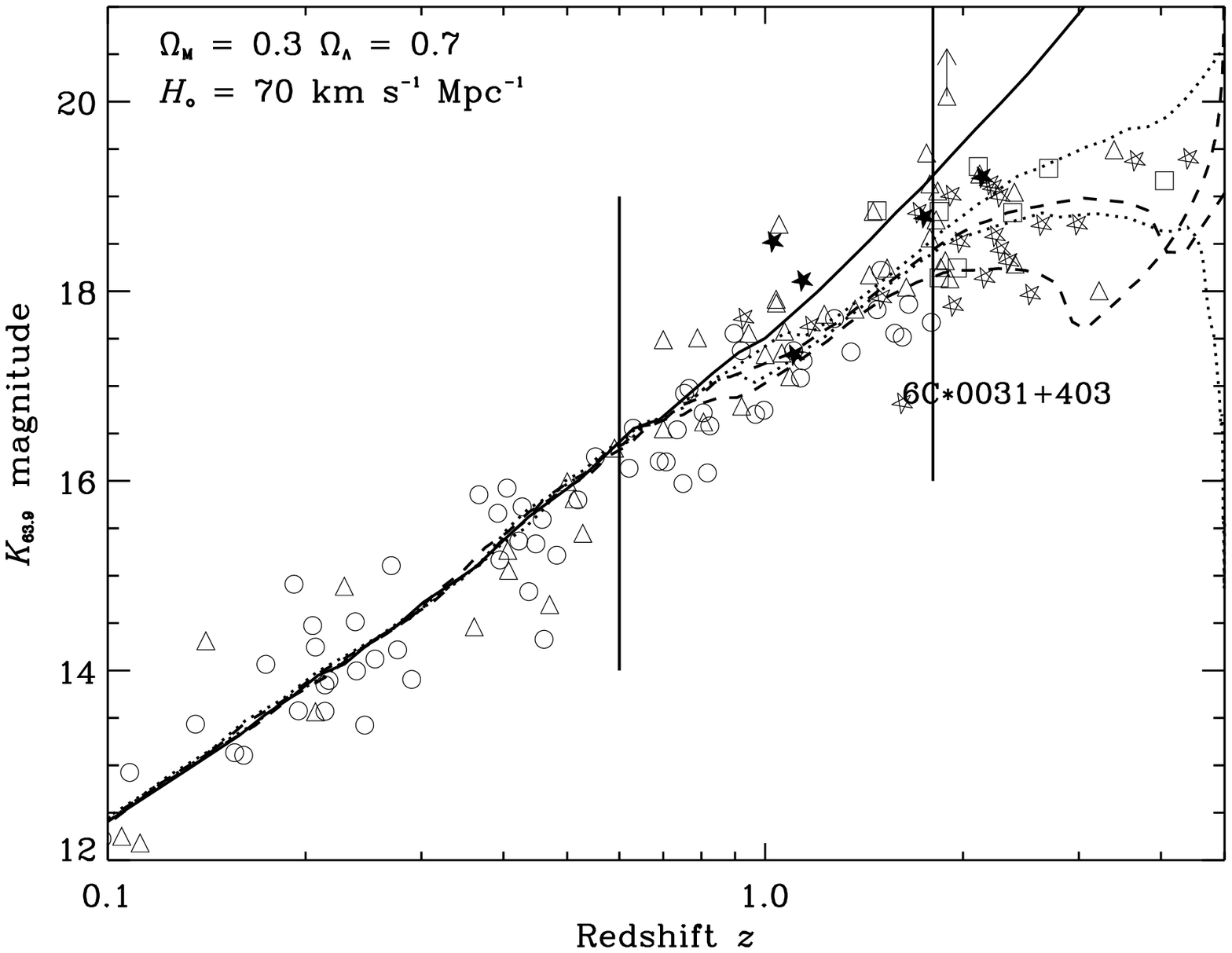} }}
{\caption{\label{fig:Kzmod} The $K-z$ Hubble diagram for 3CRR, 6CE and
6C* radio galaxies as defined in Fig~\ref{fig:kzdiag}. 
The two
vertical lines show the redshift above which the alignment effect
begins to be seen ($z > 0.6$) and the higher redshift at which we
chose to split the data ($z = 1.8$). The solid curved line is the
predicted curve for galaxies which do not evolve (as described in the text). The dashed line is
the model for a star-burst lasting 1 Gyr starting at $z = 5$ and the
dotted curve is the same evolutionary model but with the same
starburst beginning at $z = 10$. The two dotted curves represent
the models of an instantaneous (0.1 Gyr) burst beginning at $z =5$ (lower) and
$z = 10$ (upper). }}
\end{figure*}

In Fig.~\ref{fig:Kzmod} we plot the $K-z$
diagram for all of the sources in our dataset. We also show four
synthetic galaxy evolution models from the `Galaxy Isochrone Synthesis
Spectral Evolution Library' (GISSEL) of Bruzual \& Charlot (1993) and
a curve representing a galaxy which undergoes no-evolution. The GISSEL files that we have used
are ones in which there is an instantaneous burst of star formation
and one in which the burst of star formation lasts 1 Gyr, with a
Salpeter IMF with a lower mass cut-off of 0.1 M$_{\odot}$ and an upper
mass cut-off of 125 M$_{\odot}$. We use two different assumptions
about the star formation history, one in which the burst of star
formation begins at $z = 5$ and one in which the burst occurs at $z =
10$.

The no-evolution curve was constructed by taking the spectral energy
distribution template from the GISSEL library,
that was found to fit the observed spectral energy distribution of
a radio galaxy at $z = 0$, and which also reproduced the near-infrared
colours. All of the curves are normalised to pass through the
low-redshift ($z < 0.3$) points.

With our data on the 6C* sample in addition to the 6CE and 7C-III
samples we find that in a low-density universe (\costwo) the data are
predominantly brighter than the no-evolution curve and are consistent
with a passively evolving stellar population with a high-formation
redshift. If this passively evolving scenario is correct then
hierarchical growth at $z < 2.5$ is not a required ingredient.  However,
this brightening may not just be due to passive evolution of the
stellar population. Non-stellar contributions from the central AGN may
also contribute a higher fraction of light at these redshifts. All of
the studies to measure the non-stellar contribution to the \kband flux
conducted to date (e.g. Leyshon \& Eales 1998; Simpson, Rawlings \&
Lacy 1999), have concentrated on the most radio luminous 3CR sources
at $z \simeq 1$, and may have little bearing on the properties of the
high-redshift sources considered in this paper. If it turns out that
the high-redshift 6C sources have $\gtsim$\, non-stellar contamination
to those of the $z \sim 1$ 3C sources (which have the same radio
luminosity) then hierarchical build up may be necessary. Note that
\kband observations of the high-redshift 6C sources will be at shorter
wavelengths than those of the 3C sources.

However, separate arguments lead us to conclude that the dominant
factor is passive evolution of a stellar population which formed at $z\,
\gtsim \,2.5$. First, recent sub-mm observations with SCUBA have shown that the
dust masses in radio galaxies are larger at $z \simeq 3$ than in
galaxies with similar radio luminosities at lower redshift (Archibald et al. 2001). This implies that the majority of
star-formation activity in these galaxies is occurring at high
redshift. Second, the discovery of six extremely red objects at $1 < z
< 2$ in the 7C Redshift Survey (Willott, Rawlings \& Blundell 2001)
with inferred ages of a few Gyrs, implies that these objects formed the bulk of their stellar population at $z
\simeq 5$. Third, detailed modelling of the optical spectrum of the
weak radio source LBDS 53W091 at $z = 1.552$ has shown that this
object is most plausibly an old elliptical, with an inferred age of
$\gtsim 3.5$~Gyr (Dunlop et al. 1996; Spinrad et al. 1997). The
further discovery of LBDS 53W069 at $z = 1.43$, with an inferred age
of $\sim 4$~Gyr (Dunlop 1999) suggests that there exists a population
of evolved, radio weak ellipticals which formed at $z \gtsim
5$. However, controversy has arisen regarding these inferred ages with
more than one group showing that alternative population synthesis
models estimate these galaxies to be much younger, e.g. 1~-~2~Gyr
(e.g. Bruzual \& Magris 1997; Yi et al. 2000 - but see Nolan, Dunlop
\& Jimenez 2001), and this should be kept in mind.
Fourth, Peacock et al. (1998) have presented an independent
argument which corroborates the ages of these objects deduced from the
gravitational collapse redshift required for consistency with the
power spectrum which again leads to a high-formation redshift for these
objects.  Therefore, the new data on the 6C* sample presented in this
paper is consistent with the results from various other observational
studies of radio galaxies in which these radio-luminous
systems formed most of their stars at epochs corresponding to very
high redshifts ($z\, \gtsim\, 2.5$), and have undergone simple
passive stellar evolution since. Willott et al. (2001) have pointed
out that such galaxies probably undergo at least two active phases:
one, at epochs corresponding to $z \gtsim 5$, when the black hole and
stellar spheroid formed, and another, at e.g. $z \sim 2$, when powerful jet
activity is triggered, or perhaps re-triggered, by an event such as an
interaction or a merger. The small scatter in the $K-z$ relation (this
paper) and sub-mm results (Archibald et al. 2001) suggest that the
second active phase has little influence on the stellar mass of the
final elliptical galaxy

We have shown that the powerful radio galaxies in our samples are
consistent with having passively evolving stellar populations.  If we
now compare the masses of these powerful radio galaxies to the derived
value of $M_{K}^{\star}$ for nearby elliptical galaxies
[$M_{K}^{\star} = -24.3$ for $H_{\circ} = 70\,{\rm km\,s}^{-1}\,{\rm
Mpc}^{-1}$ (Kochanek et al. 2000)], we find, if passive evolution is
accounted for, that the powerful radio galaxies considered in this
paper are consistent with being $\approx 5 L^{\star}$ throughout the
redshift range $0 < z \ltsim\, 2.5$.

\section{Conclusions}\label{sec:kzconclusions}

We have obtained near-infrared images for the filtered 6C* radio
sample. With this filtered survey we are able to significantly
increase coverage of the $K-z$ Hubble diagram for radio galaxies at $z
> 2$ and have reached the following conclusions.

\begin{itemize}

\item 3CRR galaxies at $z > 0.6$ show a tighter range in near-IR
luminosity than at $z < 0.6$. This can be explained by a radio
luminosity dependent component affecting the lower luminosity host galaxies more
than the higher luminosity host galaxies, making the correlation tighter. 
Alternatively it may
be a consequence of the small range of physical conditions needed to form
the most massive black holes and hence the most luminous radio sources.
With the addition of the fainter 6CE and 6C* samples we find
that the dispersion in \kband magnitude increases with
decreasing radio luminosity, consistent
with both of the above explanations.

\item 6CE/6C* galaxies at $0.6 < z < 1.8$ have fainter \kband
magnitudes than the 3CRR galaxies at similar redshifts, implying a
link between radio luminosity and near-infrared luminosity, in
agreement with earlier work (e.g. Dunlop \& Peacock 1993; Eales et al. 1997). This
may be a consequence of a link between radio luminosity/black
hole mass and the mass of the host galaxy. The more massive galaxies
harbour the most massive black holes, in accordance with recent work
on nearby non-active galaxies (e.g. Magorrian et al. 1998).

\item We find no evidence from our emission-line corrected \kband
magnitudes for an increase in the dispersion of 6CE and 6C* radio
galaxies from $z \sim 0.6$ to $z \sim 2.5$, indicating that we are not yet
probing into the epoch  of formation in which we would expect to see a
broader range in magnitudes.

\item We find that radio galaxies are consistent with a passively
evolving population which formed at high-redshift ($z\, \gtsim\,
2.5$), in agreement with recent sub-mm observations of radio galaxies
(Archibald et al. 2001) and extremely red objects from the 7C Redshift
Survey (Willott et al. 2001) and other faint radio surveys
(e.g. Dunlop et al. 1996).

\item The host galaxies of powerful radio galaxies appear to be associated with
galaxies with a luminosity distribution with a high mean of $\approx 5
L^{\star}$ and a low-dispersion ($\sigma \sim 0.5$~mag) up to at least
$z \sim 2.5$ . 

\end{itemize}

\section*{Acknowledgements}
We thank the many people involved in the acquisition of the imaging
data on the 6C* sample, and in particular, Jim Dunlop for obtaining
the UKIRT image of 6C*0032+412 and Paul Hirst for 6C*0107+448. Hyron
Spinrad, Daniel Stern, Arjun Dey and Adam Stanford who all helped with
the Keck optical imaging observations. The United Kingdom Infrared
Telescope is operated by the Joint Astronomy Centre on behalf of the
U.K. Particle Physics and Astronomy Research Council. W.M.Keck
Observatory is operated as a scientific partnership among the
University of California, the California Institute of Technology, and
the National Aeronautics and Space Administration. The Observatory was
made possible by the generous financial support of the W.M.Keck
Foundation.

\section*{Appendix A: $I$- band imaging from the Keck
telescope}\label{app:a}
In this appendix we present our Keck $I-$ band images of 6C*0041+460
and 6C*0050+419. These images were obtained using LRIS (Oke et
al. 1995) on the Keck-II telescope using a TEK 2048 $\times$ 2048 CCD
detector with a pixel scale of 0.212 arcsec~pixel$^{-1}$. 

\begin{figure*}[!ht]
{\hbox to \textwidth{\epsfxsize=0.45\textwidth
\epsfbox{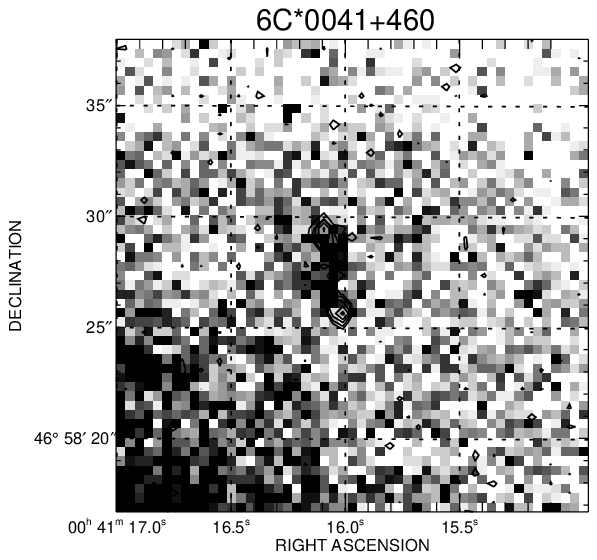} 
\epsfxsize=0.45\textwidth
\epsfbox{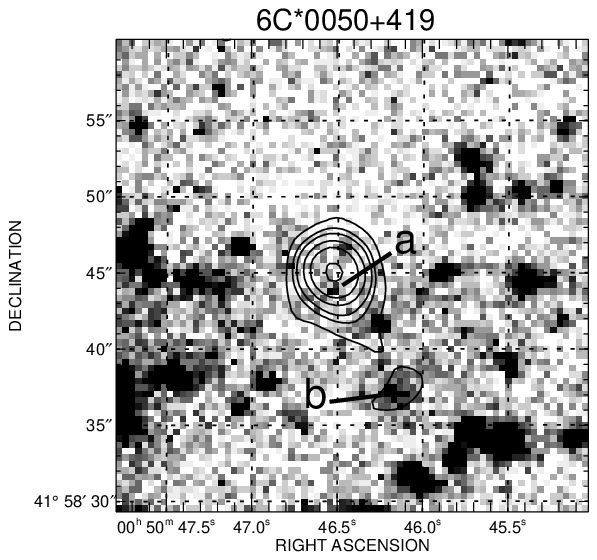} }}
{\caption{\label{fig:Iimages} \iband images of the 6C* radio sources
 6C*0041+460 (left) and 6C*0050+419 (right) taken with the Keck-II
telescope, with radio contours overlaid, the frequency of which are
given in Table~\ref{tab:kmags}. The image of 6C*0041+460
is a 300 second exposure and the ID has a magnitude of $I = 23.51$ (8
arcsec aperture) and the \iband image has been shifted to the same
astrometric frame as our \kband image (a shift of $\approx 1$~arcsec east). The image of 6C*0050+419 was a 450 second exposure and the
magnitude of the probable ID associated with the faint radio emission
to the south-west of the bright radio component (see
Sec.~\ref{sec:notes}) is $I =23.10$ (8 arcsec aperture).}}
\end{figure*}

\section*{Appendix B: Imaging data on the sources excluded from the 6C* sample}\label{app:b}
In this appendix we show \kband and \rband images of the five sources
which are now excluded from the 6C* sample as more recent radio
observations reveal that their angular sizes are larger than the
cut-off value of 15~arcsec.

\begin{figure}[!ht]
{\hbox to \textwidth{\epsfxsize=0.45\textwidth
\epsfbox{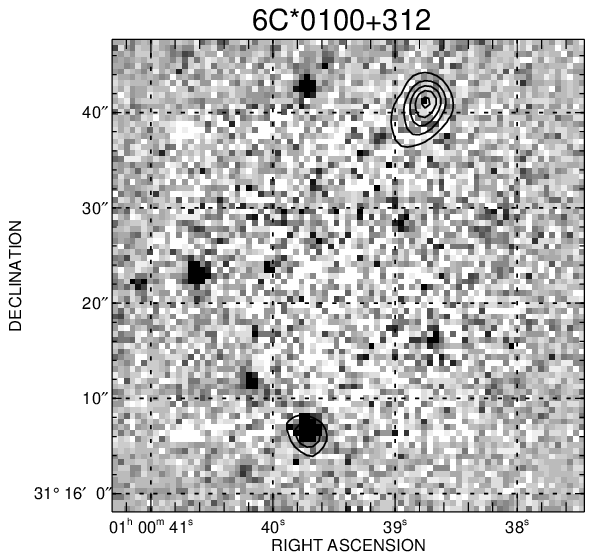}}}
{\caption{\label{fig:Kreject} \kband image of the radio source
6C*0100+312. The optical identification is unresolved and co-spatial with the core of the
radio emission. There is also another lobe of radio emission, which is
not shown on this image, to the south-east of the core at 01 00 41.20
+31 16 17.0 (B1950) which is co-linear with the core and the lobe to the
north-west.}}
\end{figure}

\begin{figure}[!ht]
{\hbox to \textwidth{\epsfxsize=0.45\textwidth
\epsfbox{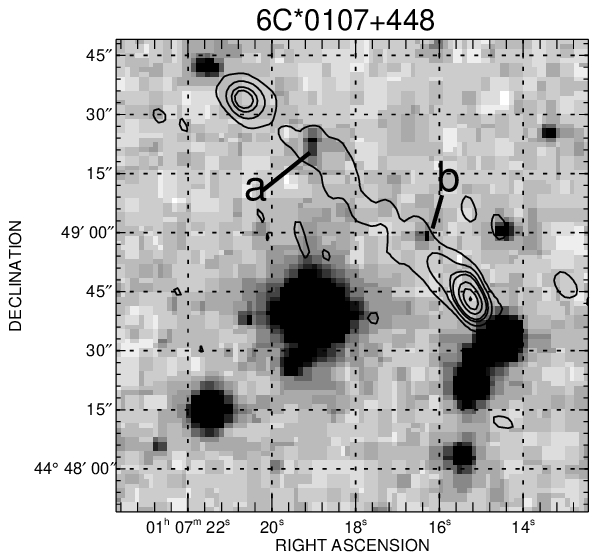}}} 
{\caption{\label{fig:kreject} $R-$band
digitised sky survey (DSS) image of 6C*0107+448 overlaid with our new 1.4
GHz radio map (peak flux contour = 26 mJy/beam and the lowest contour
= 1 mJy/beam).  There are two possible optical counterparts for this
radio source on the DSS image marked `a' and `b' on the image, with
object `a' being the most plausible ID.  Our \kband image does not
cover the area of sky containing the central region of this radio
source.}}
\end{figure}

\begin{figure}[!ht]
{\hbox to \textwidth{\epsfxsize=0.45\textwidth
\epsfbox{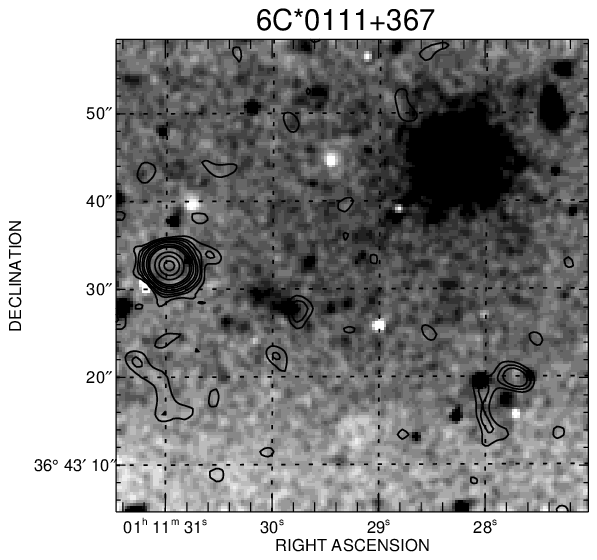}}}
{\caption{\label{fig:Rreject} WHT \rband image of 6C*0111+367.  The
optical identification is (within the astrometric uncertainties)
co-spatial with the central radio component in our 8.4 GHz map, with
the two radio lobes co-linear with each other and the core, the image
has been smoothed with a 2D Gaussian of $\sigma =1$ pixel for
presentation purposes.}}
\end{figure}

\begin{figure}[!ht]
{\hbox to \textwidth{\epsfxsize=0.45\textwidth
\epsfbox{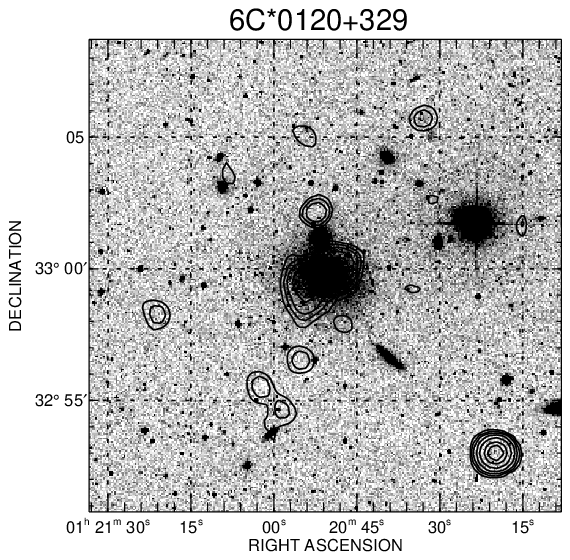}}}
{\caption{\label{fig:Rreject0120} \rband DSS image of 6C*0120+329. The
1.4 GHz NVSS (Condon et al. 1998) map shows there is a radio source
associated with an extended galaxy at $z = 0.0164$ (Jarvis et al. 2000). 
}}
\end{figure}

\begin{figure}[!ht]
{\hbox to 0.5\textwidth{\epsfxsize=0.45\textwidth
\epsfbox{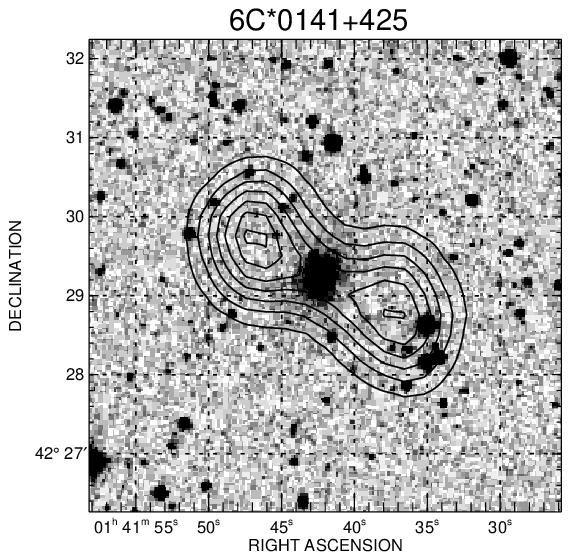}}}
{\caption{\label{fig:Rreject0141} \rband DSS
image of 6C*0141+425 overlaid with the 1.4 GHz NVSS map. One can
see the optical identification is co-spatial with the midpoint of the
radio peaks.}}
\end{figure}

\end{document}

%% file: macro.tex
\newcommand{\gtsim}{\mbox{{\raisebox{-0.4ex}{$\stackrel{>}{{\scriptstyle\sim}}
$}}}}
\newcommand{\ltsim}{\mbox{{\raisebox{-0.4ex}{$\stackrel{<}{{\scriptstyle\sim}}
$}}}}

\newcommand{\figstart}[1]
    { \begin{figure}[htb]
      \begin{picture}(0,#1) }
\newcommand{\figend}[4]
    { \end{picture}
      \special{#1}
      \caption[#2]{#3}
      \label{#4}
      \end{figure} }